\documentclass[10pt,letterpaper]{article}
\usepackage[top=0.85in,left=2.75in,footskip=0.75in]{geometry}

\usepackage{amsmath,amssymb}

\usepackage{changepage}

\usepackage{textcomp,marvosym}

\usepackage{cite}

\usepackage{nameref,hyperref}

\usepackage[right]{lineno}

\usepackage[nopatch=eqnum]{microtype}
\DisableLigatures[f]{encoding = *, family = * }

\usepackage[table]{xcolor}

\usepackage{array}

\newcolumntype{+}{!{\vrule width 2pt}}

\newlength\savedwidth

\raggedright
\usepackage[aboveskip=1pt,labelfont=bf,labelsep=period,justification=raggedright,singlelinecheck=off]{caption}

\makeatletter
\renewcommand{\@biblabel}[1]{\quad#1.}
\makeatother

\usepackage{lastpage,fancyhdr,graphicx}
\usepackage{float}
\fancyheadoffset[L]{2.25in}
\fancyfootoffset[L]{2.25in}
\usepackage{amsthm}
\usepackage{mathrsfs}
\usepackage{bbm}
\usepackage{mathtools}
\usepackage{algorithm,algpseudocode}
\usepackage{nicefrac}
\usepackage{verbatim}
\usepackage{subcaption}
\usepackage{diagbox}
\usepackage{booktabs,colortbl}
\usepackage{cleveref}
\usepackage{tikz}
\usepackage{tikz-cd}
\usepackage{enumitem}
\usepackage{xparse}
\usepackage{todonotes}
\usepackage{pgffor}
\usepackage{adjustbox}
\usepackage{longtable}

\graphicspath{{figures/}}

\setlist[itemize]{label=\tiny\textbullet}
\setlist[enumerate]{label=\normalfont(\roman*)}

\definecolor{FTChampagnePink}{HTML}{F2DFCE}

\crefname{equation}{equation}{equations}
\crefname{figure}{figure}{figures}
\crefname{section}{section}{sections}
\crefname{subsection}{section}{sections}
\crefname{lem}{lemma}{lemmas}
\crefname{ex}{example}{examples}

\tikzset{
  symbol/.style={
    draw=none,
    every to/.append style={
      edge node={node [sloped, allow upside down, auto=false]{$#1$}}
    }
  }
}

\newtheorem{definition}{Definition}

\newtheorem{example}{Example}

\newtheorem{remark}{Remark}

\NewDocumentCommand{\CF}{O{\R} O{}}{%
  \operatorname{CF}_{#2}(#1)%
}

\newcommand{\deff}[1]{\textbf{#1}}

\def\Z{{\mathbb Z}}
\def\R{{\mathbb R}}

\def\1{\mathbf{1}}
\renewcommand{\phi}{\varphi}

\def\Int{\text{Int}}

\def\ect{\text{ECT}}
\def\samp{\text{SampEuler}}

\NewDocumentCommand{\transform}{s O{\phi} d<> O{\kappa}}{%
  \IfBooleanTF{#1}{
    \operatorname{T}_{#4}%
  }{
    \operatorname{T}_{#4}\left[#2\right]%
    \IfNoValueTF{#3}{}{\left(#3\right)}%
  }%
}

\newcolumntype{L}[1]{>{\raggedright\arraybackslash}p{#1}}

\begin{document}
\vspace*{0.2in}

\begin{flushleft}
{\Large
\textbf\newline{Topological Inference for Organoids} 
}
\newline
\\
Haochen Yang\textsuperscript{1,3,4},
Byung Ho Lee\textsuperscript{4},
Anne Grapin-Botton\textsuperscript{4},
Heather A. Harrington\textsuperscript{1,3,4,5*},
Helen M. Byrne\textsuperscript{1,2*}
\\
\bigskip
\textbf{1} Mathematical Institute, University of Oxford, Oxford OX2 6GG, United Kingdom
\\
\textbf{2} Ludwig Institute for Cancer Research, Nuffield Department of Medicine, University of Oxford, Oxford OX3 7DQ, United Kingdom
\\
\textbf{3} Centre for Systems Biology Dresden, Dresden 01307, Germany
\\
\textbf{4} Max Planck Institute of Molecular Cell Biology and Genetics, Dresden 01307, Germany
\\
\textbf{5} Technische Universit\"{a}t Dresden, Dresden 01062, Germany
\\
\bigskip

* harrington@mpi-cbg.de, helen.byrne@maths.ox.ac.uk

\end{flushleft}
\section*{Abstract}

The reproducibility of organ morphology and the extent to which morphogenesis can be predicted by computational models remain difficult to quantify, particularly for organs containing complex networks of fluid-filled lumina. Topological data analysis (TDA) provides a mathematically principled framework for encoding shape through computable topological features and is therefore well suited to characterizing such systems. Here, we combine TDA, biophysical simulation, and Bayesian inference to study lumen morphogenesis in pancreatic organoids.

Lumen formation is governed by physical processes that are challenging to measure directly, including cell proliferation and luminal osmotic pressure. We simulate organoid development using a phase-field model and address the inverse problem of inferring these parameters from either time-lapse images or single morphological snapshots. Since lumen architectures vary substantially in size, structure, and connectivity, conventional geometric descriptors provide only a partial representation of their morphology. We therefore represent each organoid using SampEuler, a topological descriptor derived from the Euler Characteristic Transform (ECT). We first demonstrate that SampEuler captures morphological information encoded by the established morphometrics including the  organoid radius, lumen count and lumen area ratio, demonstrating it as a faithful and compact representation of organoid geometry.

To avoid biased estimation with insufficient statistics, we then perform parameter inference using an approximate Bayesian computation (ABC) rejection framework with a Wasserstein distance defined on the SampEuler descriptor space. Using synthetic organoids with known ground-truth parameters, our approach accurately recovers the osmotic pressure and the proliferation rate while revealing a compensatory trade-off between the two processes. Applied to experimental data from ten pancreatic organoids, the inferred posterior distributions are consistent with biological expectations. Together, these results establish a non-destructive, image-based pipeline for estimating otherwise inaccessible physical parameters governing lumen formation and highlight the potential of topological representations for linking complex biological morphology to mechanistic models.

\section*{Author summary}
Many organs contain fluid-filled lumina whose shapes are essential for tissue function. In the pancreas, lumen morphology is controlled by at least two key physical processes: luminal osmotic pressure and the rate of cell proliferation. Computational models can predict how these processes shape organoid development, but it remains challenging to infer values of the underlying physical parameters from observed morphology. We address this problem by combining a topological representation of organoid shape based on the ECT combined with Bayesian statistical inference. Our approach SampEuler quantifies complex lumen architectures and compares experimental images with simulations to identify the parameters most likely to have generated a given morphology. Using synthetic datasets, we accurately recover known parameter values, and when applied to experimental data from pancreatic organoids grown in vitro, we obtain physically realistic estimates. Since the method relies only on imaging data, it provides a fast, non-destructive tool for studying morphogenesis in multiple organs.


\section*{Introduction}

\begin{figure}[htbp]
\centering
\includegraphics[width=\textwidth]{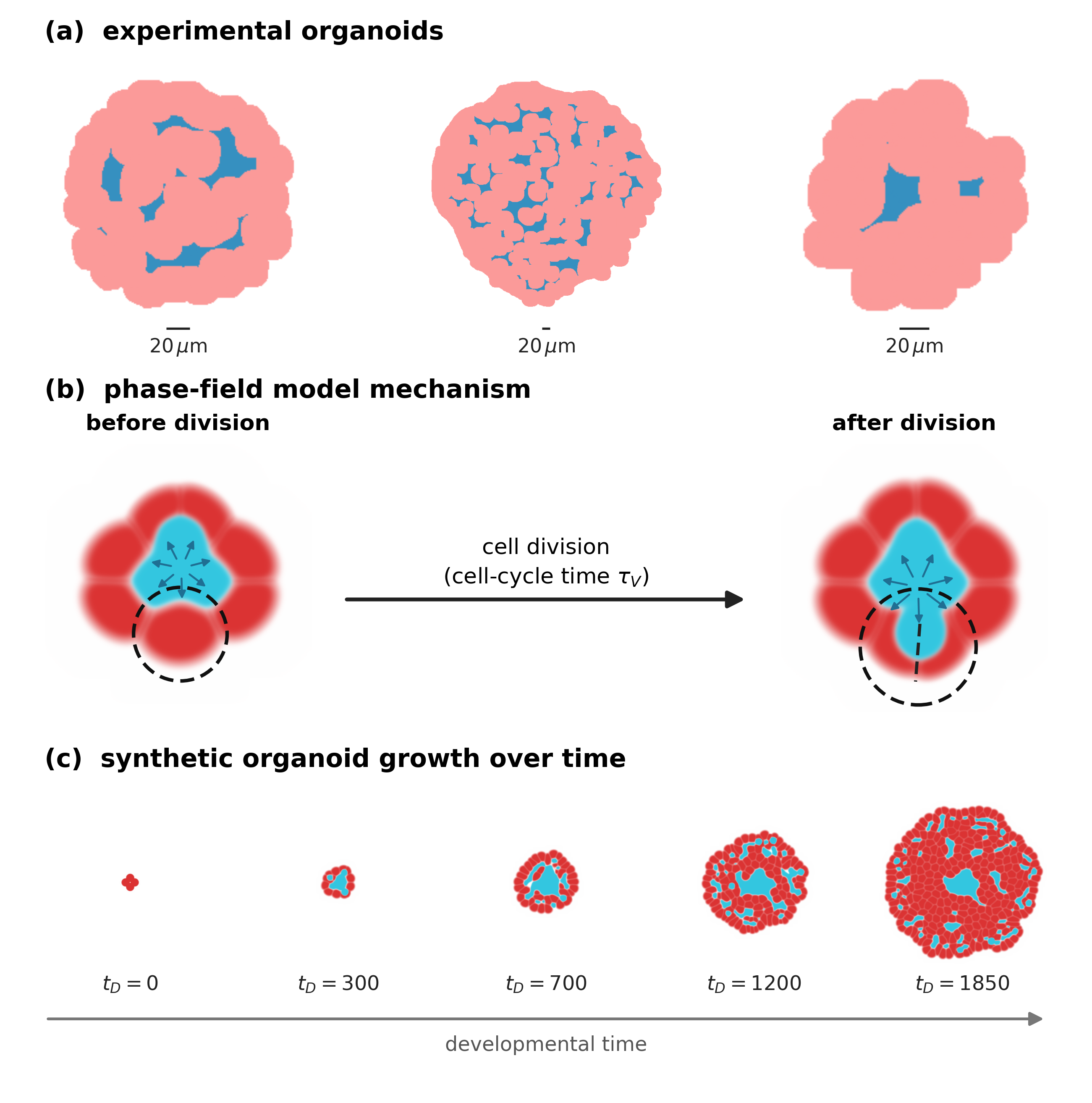}
\caption{{\bf Experimental organoids and the phase-field model.} \textbf{(a)}~Example segmented experimental organoid images, with cells in pink and lumina in blue. Each image is a snapshot of a fixed end-point. \textbf{(b)}~Cell division and lumen expansion in the phase field model, with cells in red and lumen in cyan. The cell marked by the dashed circle divides into the two daughter cells circled on the right, separating along the division plane (dashed line) and generating a small lumen at the abscission point. Arrows mark the outward osmotic pressure $\xi$ that inflates the lumen. Cells divide with a cell-cycle time $\tau_V$, with larger $\tau_V$ meaning slower division. \textbf{(c)}~A typical simulated organoid time course, generated by the phase field model.}
\label{fig:model}
\end{figure}

An organ's morphology is intimately linked to its function and understanding how organ shape emerges during development remains a central challenge in developmental biology. Organoids are self-organising three-dimensional cellular assemblies that recapitulate key features of organ architecture and have become powerful experimental systems for studying morphogenesis~\cite{lancaster2014organogenesis, clevers2016modeling}. Many organs rely on networks of fluid-filled lumina for their function, making lumen formation a fundamental step in epithelial development~\cite{sigurbjornsdottir2014molecular, lubarsky2003tube}. The pancreas provides a striking example. During development, an initially stratified epithelium generates numerous microlumina, which subsequently fuse into an interconnected luminal plexus before remodelling into a branched ductal tree~\cite{villasenor2010epithelial,bastidas2017cellular,larsen2017molecular,kesavan2009cdc42}. Pancreatic organoids reproduce key aspects of this process in vitro (\Cref{fig:model}a), providing a controlled system for investigating the physical and biological mechanisms that determine ductal architecture~\cite{greggio2013artificial}.

Recent experimental and theoretical studies have identified two major determinants of luminal morphology in pancreatic organoids: cell proliferation and luminal pressure~\cite{torres2021tissue, chan2020integration, tanida2025predicting, lee2026permeability}. Building on these observations, Tanida et al. developed a phase-field model of de novo lumen formation in pancreatic organoids~\cite{tanida2025predicting}, which was later combined with experiments to show that lumen morphology is governed primarily by the interplay between cell proliferation and osmotic pressure~\cite{lee2026permeability}. In this modelling framework, each cell and the enclosed lumen are represented as smooth phase fields on a computational grid (\Cref{fig:model}b). Cells grow and divide at a rate determined by their cell-cycle time $\tau_V$, while lumen expansion is driven by an osmotic pressure $\xi$. Starting from a small cellular aggregate, the model generates mature organoid morphologies spanning a range of lumen shapes and sizes (\Cref{fig:model}c).

The mechanistic model establishes a forward map from biological parameters to morphology. Experimentally, however, only the resulting morphology is observed (\Cref{fig:model}a) and not the underlying cell-cycle time, osmotic pressure, or developmental stage that produced it. Moreover, different parameter combinations can generate morphologies that are visually similar, making it difficult to infer biological drivers directly from conventional image measurements. Recovering the underlying parameters from morphology would allow organoids to be positioned within a physical parameter space and provide a quantitative foundation for comparing experimental conditions. This motivates the inverse problem considered here: given an observed organoid morphology, can we infer the parameters that generated it?

Solving this inverse problem is challenging because the phase-field model is stochastic, computationally expensive, and does not admit a tractable likelihood. Such settings are natural targets for likelihood-free inference methods, particularly approximate Bayesian computation (ABC), which replaces likelihood evaluation with comparisons between observed and simulated data based on summary statistics~\cite{pritchard1999, beaumont2002abc}. The quality of ABC inference depends critically on these summary statistics. Conventional morphometric descriptors, such as lumen area, perimeter, and/or lumen count, capture only limited aspects of morphology~\cite{lee2026permeability,PPR:PPR811193}. Distinct lumen architectures can share identical values of these descriptors~\cite{lee2026permeability}. Their sufficiency for the phase-field parameters is not established, so ABC based on them is not guaranteed to recover the posterior conditional on the complete observed shape.

To overcome this limitation, we employ topological data analysis (TDA), a mathematical framework that quantifies shape through topological features and has proven effective for analysing complex spatial data in biology and other scientific domains~\cite{taylor2015topological, li2025counting, prat2026dark}. There is increasing interest in using TDA to estimate the parameters of mechanistic models in spatial biology~\cite{nardini2021topological, thorne2021tabc, wenzel2025topologically, mcdonald2026topological, liu2026multi}. Here, we use SampEuler~\cite{yang2026topological}, a recently introduced shape descriptor based on the Euler characteristic transform (ECT)~\cite{turner2014persistent}. In two dimensions, the Euler characteristic is the number of connected regions minus the number of holes. The ECT records how this quantity changes as we sweep across the shape progressively along any directions. Each scanning direction produces an Euler characteristic curve. We regard two objects equal after a series of translations, rotations, and reflections as having the same shape~\cite{kendall1999shape}. After centring each organoid to remove translation, we consider the uniform distribution of these curves across all directions. This distribution, known as the ECT pushforward measure, is injective up to rotations and reflections~\cite{curry2022many}. SampEuler approximates this distribution using a finite sample of directions. This makes SampEuler distinguish two objects if and only if they have different shape while remaining stable to image perturbations~\cite{yang2026topological}. Morphologies can be compared from the Wasserstein distances between their SampEuler summaries.

In this paper, we combine SampEuler and ABC to infer key physical parameter values governing pancreatic organoid morphogenesis. From a single organoid image, we seek to recover cell-cycle time ($\tau_V$), osmotic pressure ($\xi$), and developmental time ($t_D$). We first demonstrate that SampEuler captures substantially richer morphological information than conventional descriptors and enables accurate nearest-neighbour recovery of parameters from simulated reference datasets. We then integrate SampEuler within an ABC-rejection framework to obtain full posterior distributions over parameters. Using synthetic data, we show that the approach accurately recovers ground-truth parameters and reveals inherent parameter trade-offs: for example, a younger organoid with a short cell-cycle time can closely resemble an older organoid with a longer cell cycle time. Using full developmental trajectories resolves much of this ambiguity and exposes biologically meaningful relationships between the cell-cycle time and the osmotic pressure. Finally, we apply the method to images of experimental pancreatic organoids and obtain physically realistic posterior distributions for their underlying physical parameters. Together, these results demonstrate that topology-based summaries enable the inference of the physical drivers of lumen formation directly from organoid morphology.

\section*{Materials and methods}

\subsection*{Pancreatic organoid imaging dataset}
Organoids were generated from dissected pancreatic buds collected from mouse embryos at stage e10.5, with 0.5 defined as noon of the day of dam’s vaginal plug detection. All experiments were performed in accordance with the German Animal Welfare Legislation (”Tierschutzgesetz”), as  approved by the federal state authority Landesdirektion Sachsen (license DD24-5131/451/8). ICR mice were kept in standardised specific-pathogen-free (SPF) conditions at the Biomedical Services Facility (BMS) of Max Planck Institute of Molecular Cell Biology and Genetics. The buds were dissociated using TrypLE (12604013, 290, ThermoFisher Scientific) treatment for 12 min in a 37$^{\circ}$C incubator, followed by mechanical dissociation using pulled glass capillaries (BR708707, BRAND/Merck). The cell aggregates were seeded into cold 75\% Matrigel (356231, Corning) in 8-well glass bottom plates (80826, Ibidi) and polymerisation was triggered at 37$^{\circ}$C for 10 min. To grow branching organoids, a medium composed of 25 ng/mL murine-FGF1 (450-33, Perprotech), 25 ng/mL murine-EGF (315-09, Perprotech), 2.5 U/mL Heparin (7980, Stemcell Technologies), 10 $\mu$M Y-27632 -dihydrochloride ROCK inhibitor (Y0503, Sigma Aldrich), 16 nM Phorbol-12-myristate-13-acetate (524400, Milipore), 100 ng/mL murine-FGF10(450-61, Perprotech), 500 ng/mL murine-Spondin-1 (315-32, Perprotech), 10\% Knockout serum (10828-028, Gibco), 1\% Penicillin-Streptomycin (15140-122, Sigma Aldrich) and DMEM/F12 (1:1) 1x (+) L-Glutamine (11320-033, Gibco) was used. The organoids were grown for 6 days in culture in an incubator at 37$^{\circ}$C 5\% CO2. Medium exchange was carried out every 2 days.

Pancreatic organoids were fixed with 4\% Formaldehyde (28908, ThermoFisher Scientific) PBS for 30 min at room temperature. Samples were blocked and permeabilised in 0.25\% Triton (T8787, Sigma Aldrich), 1\% bovine serum albumin (BSA; A3059, Sigma Aldrich) in PBS for 6 h at room temperature. Pancreatic organoids were incubated in a solution of 0.25\% Triton, 1\% BSA in PBS containing the primary antibodies overnight at 4$^{\circ}$C, and in secondary antibody solution in 0.25\% Triton, 1\% BSA in PBS overnight at 4$^{\circ}$C. To stain nuclei, DAPI (ab228549, Abcam) was added at 1:1000 in 0.25\% Triton, 1\% BSA in PBS for 4 hours at room temperature after the incubation of secondary antibody. The primary antibodies used to mark the lumen were anti-Ezrin (3C12) (sc-58758, Santa Cruz). Secondary antibodies used were anti-Mouse IgG H\&L (Alexa Fluor 488) preadsorbed (ab150117, Abcam), anti-Mouse IgG H\&L (Alexa Fluor 647) preadsorbed (ab150111, Abcam). Primary and secondary antibodies were used with a dilution factor of 1:400 for all immunofluorescence experiments.

Immunofluorescence samples were imaged on a ZEISS LSM 700 inverted single-photon point-scanning confocal microscope built on an Axio Observer.Z1 stand with a motorised stage. Signal was collected through two photomultiplier tubes (PMTs) alongside a transmission detector (T-PMT). Excitation was provided by laser diodes at 405, 488, 555, and 639 nm, and images were taken with Zeiss Plan-Apochromat objectives (20$\times$/0.8 NA air and 25$\times$/0.8 NA water/glycerol/oil). Acquisition was controlled by Zeiss ZEN 2012 SP5 FP3 (black) software (64-bit, version 14.0.25.201).

To segment the lumen and whole-organoid structure, images were denoised with Noise2Void~\cite{krull2019noise2void}. Epithelial-marker channels (nuclei and membranes) were summed using pyclesperanto-prototype~\cite{pyclesperanto}, then processed with a Gaussian blur (sigma xyz = $0.75$--$1.5$) and top-hat background removal (radius xyz = $20$--$30$). These channels, together with the enclosed lumen, were manually annotated in Napari to train an APOC model~\cite{apoc}, which was then applied to segment the epithelium. Prediction errors were corrected manually in Napari or semi-automatically with pyclesperanto-prototype's binary processing functions. Finally, 2D segmentations were obtained by selecting the largest-area z-plane from the 3D output for downstream analysis.

\subsection*{Phase-field organoid model}
We use a two-dimensional phase-field model of organoid development to generate synthetic organoid data \cite{tanida2025predicting,lee2026permeability}. Throughout this study the model is used with its published default parameter settings. Here we provide a brief overview and refer the reader to the original publications for a full description. The governing free-energy functional and evolution equations are given in \nameref{S2_Appendix}. 

The model represents each epithelial cell and the lumen as smooth phase fields defined on a fixed computational grid (\Cref{fig:model}b). The dynamics are governed by a shared free-energy functional, with all phase fields evolving to reduce the total free energy. Several biophysical mechanisms contribute to this energy. Cell-cell overlap increases the free energy and therefore generates a repulsive force between neighbouring cells. Increasing the contact area between adjacent cells lowers the free energy, representing cell-cell adhesion, while shortening the cell boundary also lowers the free energy, promoting smooth cell shapes. Additional terms drive cells towards a prescribed target volume and promote lumen expansion through osmotic pressure, causing the lumen to inflate against the surrounding epithelial layer. 

Cell proliferation is incorporated via a volume-dependent division rule. A cell divides once its volume exceeds a prescribed threshold, producing two daughter cells separated by a division plane determined by a spindle-pole model~\cite{tanida2025predicting}. A new microlumen is seeded between the daughter cells at division. Simulations are initialised with four circular cells of radius $0.9$, centred at offsets $(\pm1,0)$ and $(0,\pm1)$ from the centre of the square domain $\Omega=[0,40.96]^2$, discretised with grid spacing $0.02$. All lengths are dimensionless, and `volume' denotes area in this two-dimensional model. In what follows, we focus on two parameters that control organoid development: the cell-cycle timescale, $\tau_V$, which determines the rate of cell growth and division, and the lumen osmotic-pressure parameter, $\xi$, which controls the strength of lumen inflation. Their roles in the governing equations are described in \nameref{S2_Appendix}; all other parameters are fixed at the default values reported in \cite{tanida2025predicting}. 

The model is dimensionless, so developmental time and organoid size are expressed in simulation units rather than physical units. Simulations are integrated from the initial condition, and a snapshot is recorded every 10 dimensionless time steps. Each simulation terminates either when it reaches a maximum duration of $5\times10^6$ model time units or when lumen leakage occurs. Leakage corresponds to rupture of the epithelial layer, allowing lumen contents to escape into the external medium. The simulation is stopped once the leaked volume exceeds a fixed threshold. Since both organoid growth and the time to leakage depend on $(\tau_V,\xi)$, the total developmental time reached by a simulation varies across parameter space.

\subsection*{SampEuler descriptor}

To quantify and compare organoid morphologies, we use SampEuler as our shape descriptor and the Wasserstein distance as our metric~\cite{yang2026topological}. We explain the construction using a worked example below. Formal mathematical definitions are provided in \nameref{S1_Appendix}.

For our organoid images, we represent each pixel as a square, together with its bounding edges and corner vertices. These elements form a cubical complex. To distinguish background, cells and lumen, we assign their squares weights of $0$, $1$ and $2$, respectively. Each edge or vertex inherits the largest weight of its incident squares. For an unweighted 2D object, the Euler characteristic is the number of connected components minus the number of holes.For a weighted 2D object, the weighted Euler characteristic is the sum of vertex weights minus edge weights plus square weights.

For a direction $\mathbf{v}$, we move a line perpendicular to $\mathbf{v}$ across the image. The sweep filtration parameter $t$ specifies the line's position along this direction, with $t=0$ when the line passes through the origin. At each $t$, we retain every vertex, edge and square lying entirely on the swept side of the line. Recording the weighted Euler characteristic of the retained parts against $t$ produces an Euler characteristic curve (ECC). The curve reflects changes in connectivity and holes as different parts of the image enter the sweep. The Euler characteristic transform (ECT) records the correspondence between directions $\mathbf{v}$ and output curves.

SampEuler samples $n$ directions uniformly around the unit circle and assigns probability mass $1/n$ to each resulting curve. This empirical distribution forgets the specific orientation of the input object and hence only record shape information. We compare two such distributions using the Wasserstein distance. The worked example below illustrates this construction.

\begin{example}
\label{ex:sampeuler}
\normalfont
We illustrate the computation of SampEuler using the simulated organoid in \Cref{fig:sampeuler_complex}A. The same procedure applies to experimental organoid images. We select the final frame and discretise it into background, cell and lumen regions (\Cref{fig:sampeuler_complex}B). We build a weighted cubical complex based on the discretised image. \Cref{fig:sampeuler_complex}C enlarges the boxed region to show the cubical complex structure.

\begin{figure}[H]
\centering
\includegraphics[width=\textwidth]{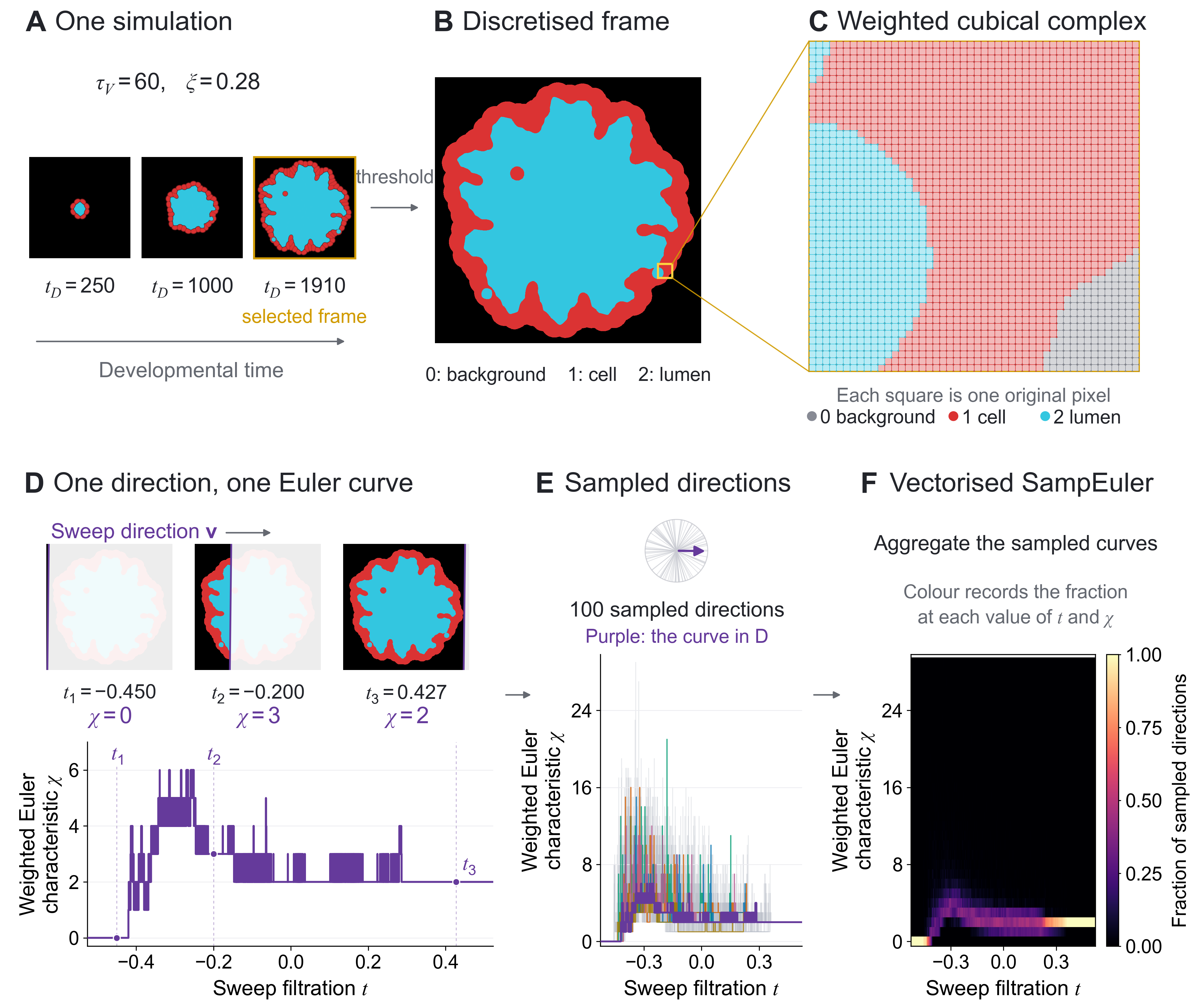}
\caption{\textbf{Computation of SampEuler for weighted cubical complexes.} \textbf{(A)}~One simulated organoid at three developmental times, shown on a common spatial scale. \textbf{(B)}~Discretisation assigns weights $0$ to background, $1$ to cells and $2$ to lumen. \textbf{(C)}~An enlargement of the boxed region in B. Each square is one original pixel. Shared edges and vertices are assigned the maximum weight of their incident squares. \textbf{(D)}~A sweep in direction $\mathbf{v}$ and its weighted Euler characteristic curve. Pale regions lie outside the swept half-plane. \textbf{(E)}~Collection of $100$ Euler characteristic curves of uniformly sampled directions forming the SampEuler empirical measure. \textbf{(F)}~Vectorisation of the SampEuler in E. At each sampled threshold $t$, colour gives the fraction of curves with each weighted Euler characteristic $\chi$.}
\label{fig:sampeuler_complex}
\end{figure}

We fix a sampled direction $\mathbf{v}$ close to horizontal. The sweep filtration parameter $t$ specifies the height threshold of the sweep. At each $t$, we retain the vertices, edges and squares lying entirely in the half-plane $\{\mathbf{x}:\langle\mathbf{v},\mathbf{x}\rangle\leq t\}$. The weighted Euler characteristic is the sum of vertex weights minus edge weights plus square weights.

\Cref{fig:sampeuler_complex}D shows three stages of this sweep and the corresponding Euler characteristic curve. At $t_1=-0.45$, the swept region contains no organoid pixels. The weighted Euler characteristic is zero for the empty set. At $t_2=-0.2$, the swept lumen has two connected components and no holes, contributing $2(2-0)=4$. The swept cell region, including its boundary, has one connected component and one hole. Its shared interface with the lumen has two connected components and one loop, giving Euler characteristic $2-1=1$. The interface edges and vertices are already included in the lumen contribution and carry weight $2$. The remaining weight-$1$ component contribution is therefore $(1-1)-(2-1)=-1$, giving a total weighted Euler characteristic of $4-1=3$. At $t_3=0.427$, the whole organoid is included in the swept half-plane, giving weighted Euler characteristic $\chi = 2$.

Repeating the calculation for $100$ uniformly sampled directions gives the collection of curves in \Cref{fig:sampeuler_complex}E and the SampEuler empirical measure is formed using the corresponding Dirac measures. For visualisation, \Cref{fig:sampeuler_complex}F records the fraction of these curves taking each integer value of $\chi$ at each of the $1000$ filtration intervals. Here, the displayed values range from $0$ to $29$. Flattening this discretised map gives a feature vector with $30{,}000$ entries.
\end{example}

\subsection*{Likelihood-free inference}

We treat the phase-field model described above as a forward model and seek to infer the parameters $\tau_V$ and $\xi$, together with the developmental time $t_D$ at which an observed organoid image was acquired. In a Bayesian framework this requires the likelihood function $p(y\mid\theta)$, where $\theta=(\tau_V,\xi,t_D)$ and $y$ denotes the observed organoid. However, the phase-field simulator is both stochastic and computationally expensive, making the likelihood analytically intractable and prohibitively costly to estimate directly by repeated simulation.

Approximate Bayesian Computation (ABC) \cite{pritchard1999,beaumont2002abc} circumvents this difficulty by replacing likelihood evaluation with simulation-based comparisons between observed and synthetic data. Given a summary statistic $S$, a distance metric $d$, and a tolerance $\epsilon$, ABC targets the approximate posterior 
\[ p\!\left(\theta \mid d(S(y),S(x)) \leq \epsilon\right), \] 
where $x$ denotes data simulated under parameters $\theta$. As $\epsilon \to 0$, this posterior converges to $p(\theta\mid S(y))$; if $S$ is sufficient, it converges to the exact posterior $p(\theta\mid y)$. Since SampEuler is a sufficient descriptor of shape, this theoretical guarantee applies directly in our setting. A range of algorithms have been developed to improve the efficiency of ABC in computationally expensive simulation models, including ABC-SMC \cite{toni2009}, Approximate Approximate Bayesian Computation (AABC) \cite{buzbas2015aabc}, and Bayesian Optimisation for Likelihood-Free Inference (BOLFI) \cite{gutmann2016bolfi}. In this work, we compare these approaches on the organoid inference problem, focusing on their computational cost, convergence properties, and practical performance.

\subsubsection*{Method selection}
The phase-field organoid model of~\cite{tanida2025predicting} constitutes the main computational bottleneck in our inference pipeline. We therefore require an inference method that (i) minimises the number of simulator evaluations, (ii) retains asymptotic convergence to the true posterior distribution, and (iii) is compatible with the SampEuler--Wasserstein distance used to quantify morphological differences between organoids. \Cref{tab:inference-methods} compares the candidate approaches in terms of their assumptions, theoretical guarantees, and computational demands in the context of our problem.


\begin{table}[htbp]
\centering
\scriptsize
\setlength{\tabcolsep}{3pt}
\renewcommand{\arraystretch}{1.25}
\begin{tabular}{|>{\raggedright\arraybackslash}p{2.3cm}|>{\raggedright\arraybackslash}p{3.2cm}|>{\raggedright\arraybackslash}p{3.4cm}|>{\raggedright\arraybackslash}p{2.2cm}|}
\hline
\textbf{Method} &
\textbf{Assumptions on the descriptor} &
\textbf{Posterior-convergence theorem} &
\textbf{Estimated wall time$^{*}$} \\
\hline
ABC-SMC \cite{toni2009,delmoral2012abcsmc} &
Distance metric only; any SampEuler form. &
\textbf{Yes} ($\epsilon\to 0$, $N\to\infty$) \cite{beaumont2009adaptive} &
$\sim 12$--$18$~months \\
\hline
ABC rejection \cite{pritchard1999,beaumont2002abc,thorne2021tabc} &
Distance metric only; any SampEuler form. &
\textbf{Yes} ($\epsilon\to 0$, $N\to\infty$) \cite{beaumont2002abc} &
$\sim 7$--$10$~days \\
\hline
AABC \cite{buzbas2015aabc} &
Euclidean features; smooth $\theta\mapsto$ features; Gaussian feature noise. &
\textbf{Yes} (training-set size and $N\to\infty$) \cite{buzbas2015aabc} &
$\sim 7$--$10$~days \\
\hline
BOLFI / parallel-GP-SL \cite{gutmann2016bolfi,jarvenpaa2021parallel} &
Any distance metric; Gaussian noise on $\log d(\theta)$; any SampEuler form. &
\textbf{No}; only $(1-1/e)$ acquisition bound \cite[\S6.3]{jarvenpaa2021parallel} &
$\sim 14$--$21$~days \\
\hline
\end{tabular}

\smallskip
\begin{minipage}{12cm}
\scriptsize\raggedright
$^{*}$Wall-time figures are estimated using a compute system
($\sim 250$~cores) and a mean per-simulation cost of $\sim 35$~h
(estimated from completed runs), multiplied by each method's approximate
required number of simulator calls and divided by the available
parallelism.
\end{minipage}

\caption{Comparison of the inference methods considered for our parameter-recovery problem. The estimated wall times correspond to the complete set of simulations each method would require, run across ten departmental machines (see the code repository for details). The parameters $\epsilon$ and $N$ in the posterior convergence theorems refer to the acceptance threshold and number of simulations, respectively.}
\label{tab:inference-methods}
\end{table}

As shown in \Cref{tab:inference-methods}, ABC-SMC is computationally prohibitive in this setting because inference must be performed sequentially over multiple generations for each target. BOLFI does not provide a posterior-convergence guarantee, and its active learning strategy must be executed separately for each target, preventing information sharing across observations. AABC is more efficient than both ABC-SMC and BOLFI, but it relies upon restrictive assumptions about the form of the summary statistics and noise model that are not satisfied by the SampEuler descriptor. In contrast, ABC rejection retains the finite-$\epsilon$ convergence guarantee, operates directly on the raw descriptor and allows a single pool of simulations to be reused across multiple targets. For these reasons, and following~\cite{thorne2021tabc}, we adopt ABC rejection as our inference method.

\subsubsection*{ABC-rejection pipeline}

\begin{figure}[htbp]
\begin{adjustwidth}{-1.6in}{-0.1in}
\centering
\includegraphics[width=\linewidth]{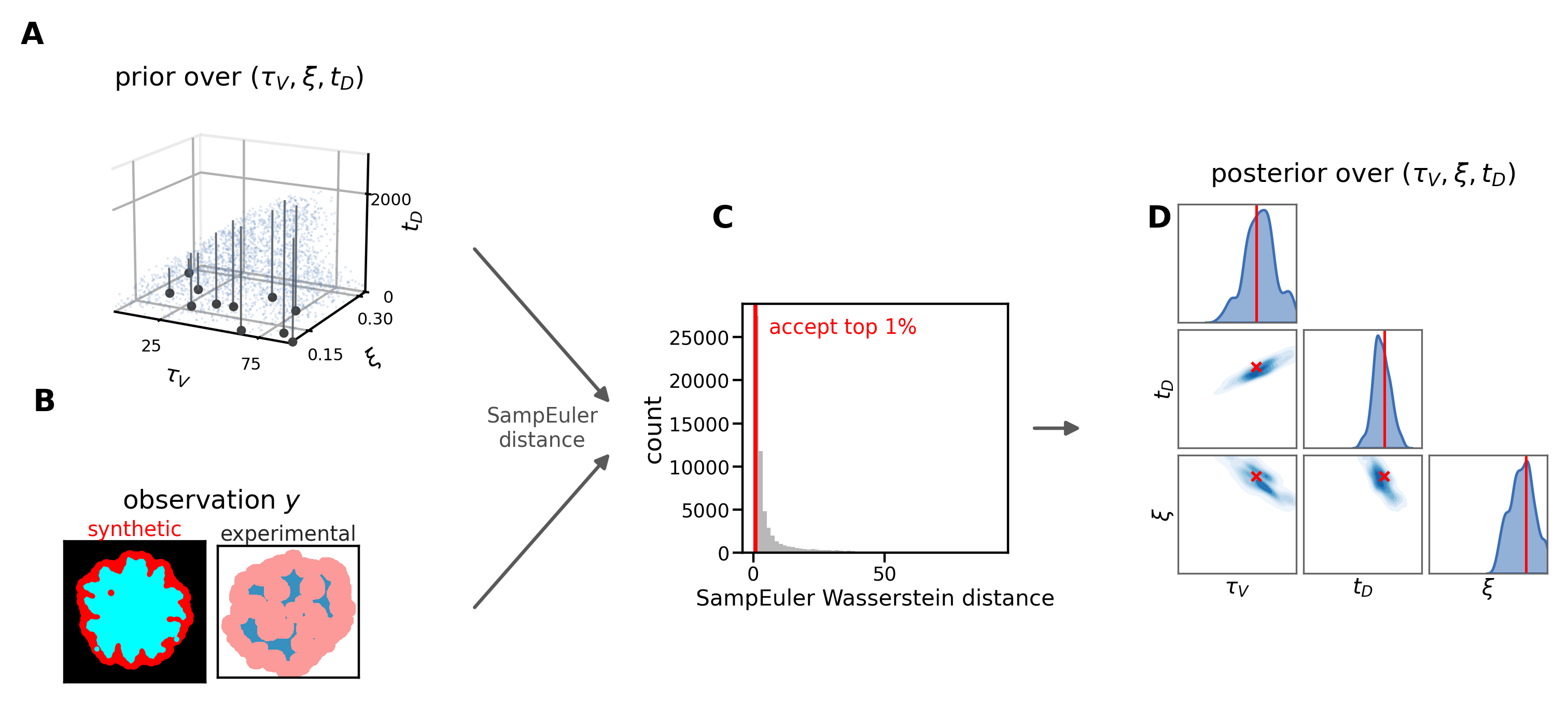}
\caption[The ABC-rejection inference pipeline.]{\textbf{The ABC-rejection inference pipeline.} \textbf{(A)}~The prior over $(\tau_V,\xi,t_D)$. Each dark point on the base plane is a parameter pair drawn from $\mathrm{Uniform}(B)$, and the line rising from it shows the time that run lasted. Pooling every frame of every run, the pale cloud, gives a uniform sample over the reachable region. \textbf{(B)}~The observation, either a synthetic organoid or a real experimental micrograph. Both enter the pipeline in the same way, each represented by its SampEuler descriptor as in \Cref{fig:sampeuler_complex}. \textbf{(C)}~The SampEuler Wasserstein distance from the observation to every candidate. The closest $1\%$ are accepted to form the posterior. \textbf{(D)}~The accepted candidates are approximate posterior samples of $(\tau_V,\xi,t_D)$, shown as a corner plot of the pairwise densities and single-parameter marginals. The distances and posteriors are those of the synthetic organoid, whose ground truth is marked in red.}
\label{fig:pipeline}
\end{adjustwidth}
\end{figure}

We infer the joint posterior over the cell-cycle time $\tau_V$, the lumen osmotic pressure $\xi$, and the developmental time $t_D$ at which the organoid image was captured. Because each experimental image is a single snapshot of unknown developmental age, $t_D$ is an inference target on the same footing as the two physical parameters. Rejection ABC needs only a forward simulator and a distance, and places no smoothness or noise-model assumptions on the descriptor; this is what enables direct comparisons of organoids with the SampEuler descriptor under the Wasserstein distance $d$~(\Cref{tab:inference-methods}). As illustrated in \Cref{fig:pipeline}, we draw $N$ parameter pairs $(\tau_V,\xi)$ from a uniform prior $\pi(\tau_V,\xi)=\mathrm{Uniform}(B)$ on the box $B=[1,90]\times[0.10,0.32]$, simulate each to its stopping time, and collect every output frame into a single global pool of $(\tau_V,\xi,t_D)$ candidates. We then uniformly sample from this pool to obtain a uniform distribution over the reachable region.

\begin{remark}[Uniform sampling over the reachable region]
Because each run stops at a parameter-dependent time $T(\tau_V,\xi)$, the reachable triples form not a cube but a region $\Omega=\{(\tau_V,\xi,t_D):(\tau_V,\xi)\in B,\ 0\le t_D\le T(\tau_V,\xi)\}$ beneath the stopping-time surface, where $B$ is the square region defined above. The simulator writes a frame every $10$ model-time units in every run, so a longer run contributes proportionally more frames. Drawing $(\tau_V,\xi)$ uniformly on $B$ and then drawing uniformly from the pooled frames of all runs therefore yields a uniform sample over $\Omega$, the standard construction for sampling uniformly from a region of varying height~\cite{robert2004monte}.
\end{remark}

For each observation $y$, we evaluate the SampEuler Wasserstein distance $d$ between $y$ and every candidate $x$ in the pool, and accept the closest $1\%$ as approximate posterior samples of $(\tau_V,\xi,t_D)$. This fixed acceptance ratio sets the tolerance $\epsilon$ to the $1\%$ quantile of the observation's distances, giving $600$ posterior samples from the pool of $60{,}000$ featurised candidates (see Implementation). Since the pool is built once and reused, the expensive simulation cost is amortised across all observations instead of being repeated for each.

\subsubsection*{Implementation}
We run the phase field organoid model of~\cite{tanida2025predicting, lee2026permeability} at its default settings (see~\cite{lee2026permeability} for details); each evaluation integrates one $(\tau_V,\xi)$ pair forward and outputs a frame every $10$ time-steps until its stopping criterion is met. Since the prior draws are independent, we parallelise the $3{,}000$ simulations across compute nodes. The SampEuler descriptor is computed for a uniform sample of $60{,}000$ frames drawn from the pool ($\approx 15\%$ of the total frames the grid produces): each shape is summarised by its Euler characteristic transform along $100$ sampled directions, evaluated at $600$ filtration values spanning $[-1.5,\,1.5]$, and two shapes are compared by the Wasserstein-$1$ distance between these curves. Since the simulator output does not have a physical scale, we bring them to a common scale by matching the diameters of the organoid's minimum-enclosing-circle, measured in pixels. Synthetic frames are produced at the simulator's canonical scale and featurised directly, whereas each experimental image is rescaled so that its organoid diameter matches the simulator's before featurisation. All code, configuration, and random seeds are available at \url{https://github.com/reddevil0623/organoid_inference}.

\section*{Results}

\subsection*{Descriptor regression on classical morphometrics}

To assess the potential of SampEuler as a shape descriptor, we test whether a simple linear model can recover classical morphometric measurements from it. Although SampEuler is naturally compared using a Wasserstein distance, ridge regression requires an $L^2$-based feature representation, so we use the vectorised form of the descriptor. We predict three morphometric quantities reported in \cite{lee2026permeability}: organoid radius (the smallest circle that contains the organoid), lumen count (the number of connected components of lumen in organoid images), and lumen area ratio (the total lumen pixel area divided by the total organoid pixel area) (\Cref{tab:morphreg}). Despite the simplicity of the linear model, predictive performance is high, achieving pooled values of $R^2=0.991$ for radius, $R^2=0.892$ for lumen count, and $R^2=0.740$ for lumen area ratio. Mean absolute errors are substantially lower than those of a baseline that predicts the dataset mean of each morphometric for every image. The reduced performance for lumen area ratio suggests a noisier or more nonlinear relationship that may be captured by more flexible models. Nevertheless, the ability of a simple linear readout to recover these hand-engineered morphometric quantities demonstrates that SampEuler encodes the information contained in standard shape descriptors, while providing a richer representation of organoid morphology.

\begin{table}[htbp]
\centering
\caption{{\bf Ridge regression of classical morphometric scalars on the vectorised SampEuler descriptor.}}
\begin{tabular}{lrrr}
\hline
Morphometric & $R^2$ & MAE & Baseline MAE \\ \hline
Organoid radius (px) & 0.991 & 3.40 & 102.4 \\
Lumen count & 0.892 & 0.82 & 5.63 \\
Lumen area ratio & 0.740 & 0.032 & 0.126 \\ \hline
\end{tabular}
\begin{flushleft}
Cross-validated performance of a ridge readout (\texttt{StandardScaler} $\to$ \texttt{RidgeCV}, $\alpha = 100$ chosen by inner leave-one-out CV) mapping the flattened vectorised SampEuler descriptor onto each scalar, across $3{,}040$ sweep frames spanning $120$ $(\tau_V, \xi)$ cells. $R^2$ and mean absolute error (MAE) are pooled over the out-of-fold predictions of grouped $5$-fold cross-validation, with folds split by $(\tau_V, \xi)$ cell so that no parameter combination appears in both training and test. The baseline is the MAE of a constant mean predictor (the mean absolute deviation of each target about its mean). Unregularised least squares overfits badly (pooled $R^2 \ll 0$ for all three targets), so the ridge penalty is essential. Organoid radius is in pixels; lumen count and lumen area ratio are dimensionless.
\end{flushleft}
\label{tab:morphreg}
\end{table}

\subsection*{Frame-level parameter recovery}
To evaluate parameter recoverability, we generate three independent synthetic test simulations and sample 20 images uniformly in developmental time from each run, giving 60 test images in total. For each image, we identify the nearest reference frame under the Wasserstein distance and compare the corresponding $(\tau_V,\xi,t_D)$ values with the ground truth.

\begin{figure}[htbp]
\centering
\includegraphics[width=\textwidth]{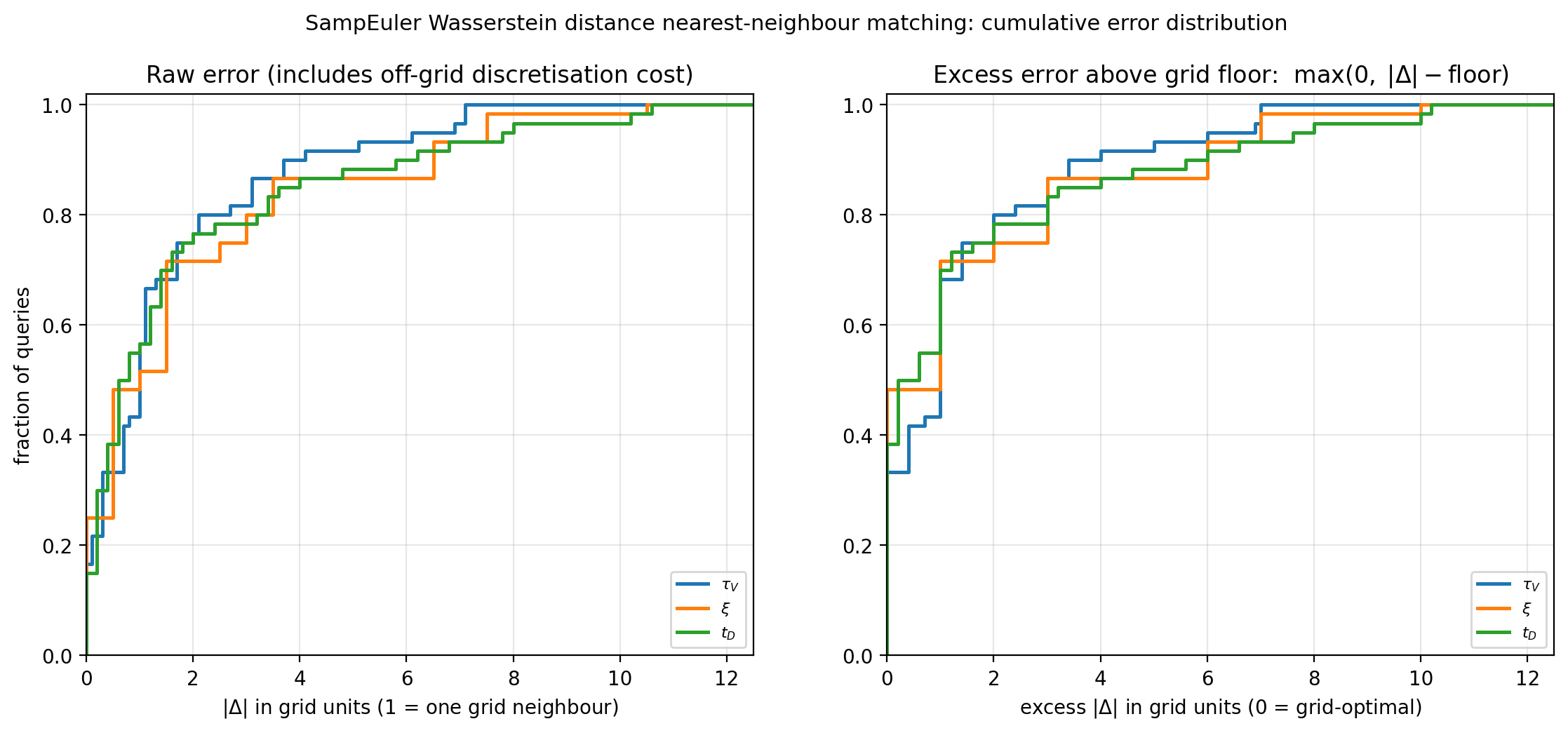}
\caption[Cumulative error distribution of the frame-level match]{\textbf{Cumulative distribution of matching errors. } The per-sample absolute grid error $|\Delta|$ for each parameter,
across $N=60$ sample frames from three synthetic test simulations; $|\Delta|=1$
corresponds to a one-grid-neighbour offset on the parameter sweep grid. \textbf{Left:}
raw error. \textbf{Right:} excess error$=\max(0,\,|\Delta|-\mathrm{floor})$, where the floor is the minimum achievable error for each sample, positive because the synthetic truths lie off the sweep grid.}
\label{fig:recovery_summary}
\end{figure}

The left panel of \Cref{fig:recovery_summary} shows the cumulative distribution of the absolute grid error $|\Delta|$ for each parameter, where $|\Delta|=1$ corresponds to a displacement of one grid neighbour. Within one grid step, the nearest-neighbour matcher recovers $57\%$ for $\tau_V$, $52\%$ for $\xi$, and $57\%$ for $t_D$. 

However, the test simulations are generated at off-grid parameter values, so even a perfect nearest-neighbour match generally incurs a non-zero error. To account for this, the right panel plots the excess error,
\[
\max(0, |\Delta| - \mathrm{floor}),
\]
where $\mathrm{floor}$ denotes the minimum achievable grid error for a given sample. Thus, $x=0$ corresponds to a grid-optimal match. Under this adjusted measure, the proportion of samples recovered within one grid step increases to $68\%$ for $\tau_V$, $72\%$ for $\xi$, and $70\%$ for $t_D$. These results demonstrate that the Wasserstein distance on SampEuler places organoids close in descriptor space when they are also close in parameter space, supporting its use for simulation-based inference.

\begin{figure}[htbp]
\centering
\includegraphics[width=\textwidth]{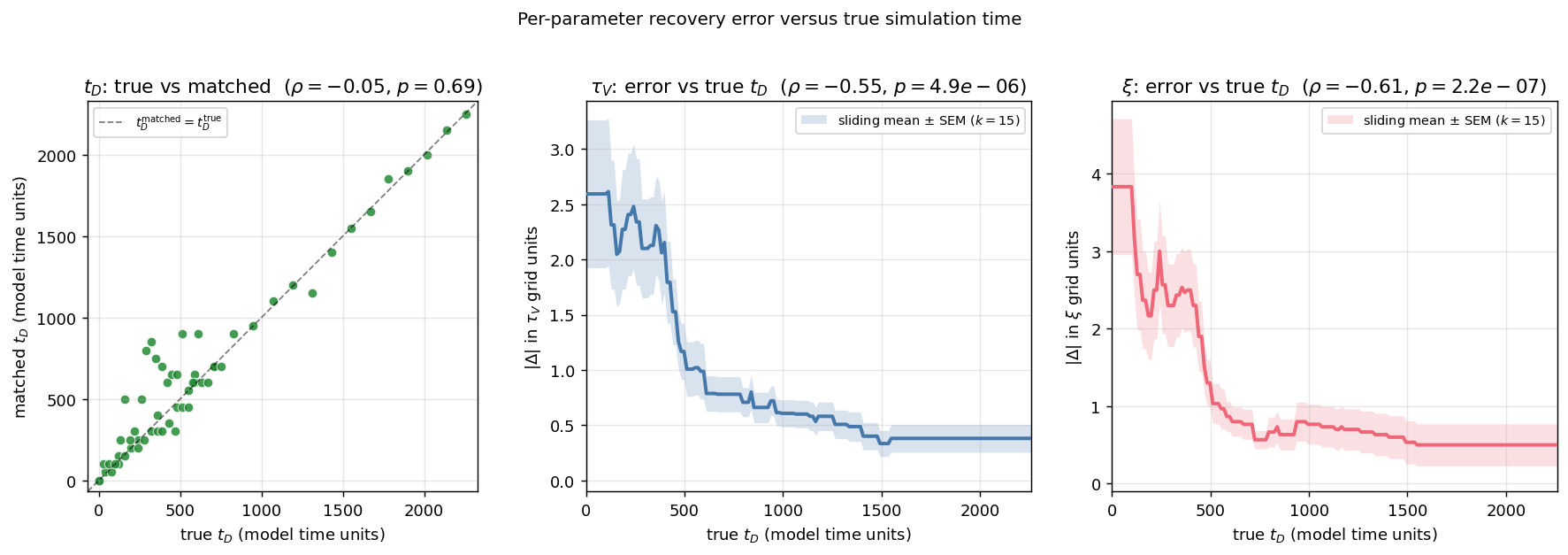}
\caption[Per-parameter recovery error versus true simulation time.]{\textbf{Per-parameter recovery error versus true simulation time $t_D$}, for $N=60$ sample frames from three synthetic test simulations. \textbf{Left:} matched $t_D$ versus true $t_D$ in model-time units, with the $y=x$ diagonal as reference. \textbf{Middle and right:} per-sample absolute grid error $|\Delta|$ for $\tau_V$ and $\xi$, summarised by a sliding-window average: at a given time we average the raw $|\Delta|$ of the $k=15$ samples nearest in true $t_D$, shading $\pm 1$ standard error of those values. The curve is traced by evaluating this average at 160 points spaced uniformly along the time axis. Spearman correlation $\rho$ and $p$-value between true $t_D$ and the per-parameter error are shown in each panel title.}
\label{fig:recovery_per_param}
\end{figure}

To understand where the matched pair misses come from, we break down the per-sample error against the true simulation time $t_D$ separately for each parameter (Fig.~\ref{fig:recovery_per_param}). For time $t_D$ (left panel) the matched and true values track the $y=x$ diagonal across the full simulation horizon, with no monotonic dependence of the absolute matching error $|\Delta t_D|$ on true $t_D$ (Spearman $\rho=-0.05$, $p=0.69$): developmental time is consistently recovered with similar accuracy at every stage. By contrast, for $\tau_V$ and $\xi$ (middle and right panels), the sliding-window mean error decreases markedly with the true simulation time: most large misses concentrate at small $t_D$, while late-time samples are typically recovered to within one grid neighbour. The trend is significant by the Spearman correlation between true $t_D$ and absolute error ($\rho=-0.55$, $p=4.9\times 10^{-6}$ for $\tau_V$; $\rho=-0.61$, $p=2.2\times 10^{-7}$ for $\xi$).

\begin{figure}[htbp]
\centering
\includegraphics[width=\textwidth]{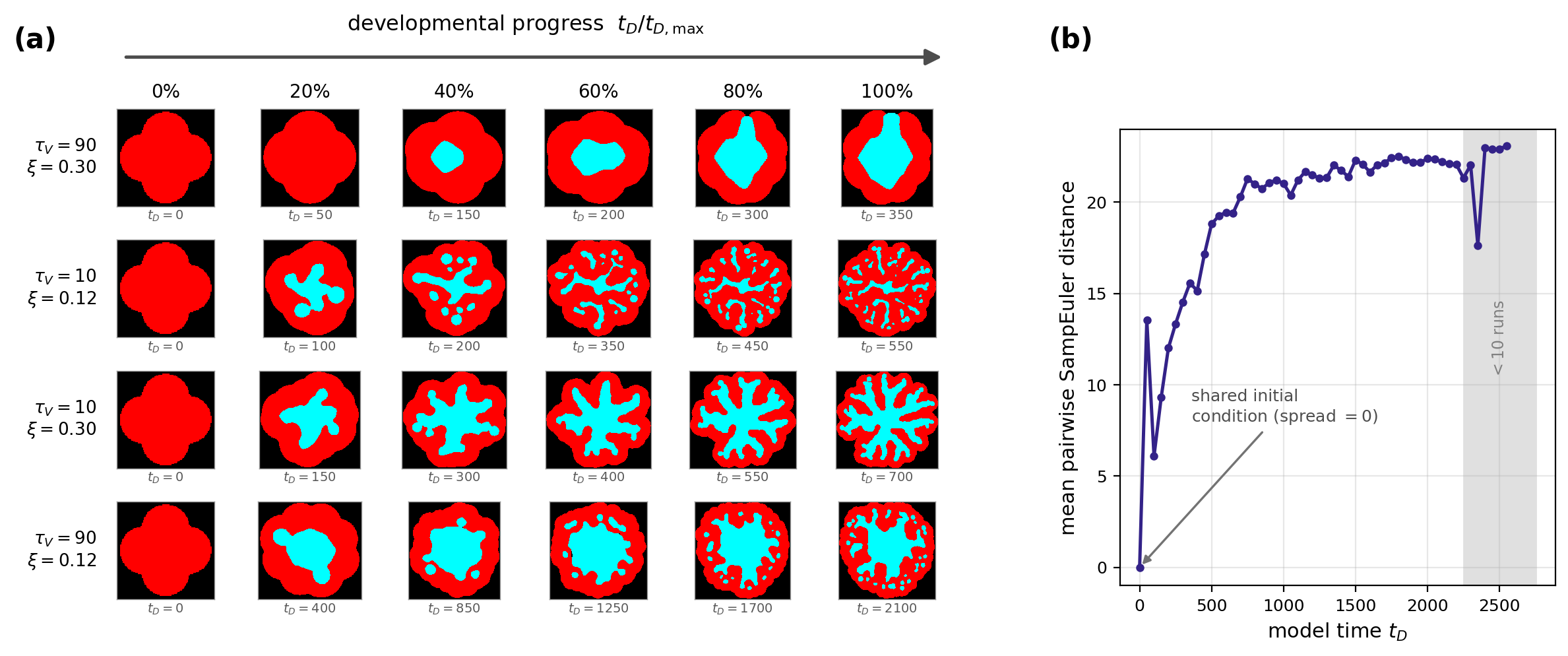}
\caption[Simulated organoids diverge from a shared initial condition.]{\textbf{Organoid morphology divergence.} Organoid morphologies are initially identical and diverge into parameter-specific shapes as they develop. \textbf{(a)} The four corners of the $(\tau_V,\xi)$ sweep grid (rows, ordered by run lifetime), each sampled at a fixed fraction of its own development (columns; the absolute model time $t_D$ is printed beneath every panel). At $0\%$ (the shared initial condition) all runs are identical; by maturity they diverge into clearly distinct fates, from fine lumen labyrinths and dendrites to large branched and single compact lumens. \textbf{(b)} The same divergence quantified over the full $120$-point sweep against absolute model time $t_D$: the spread is the mean pairwise distance between the vectorised SampEuler descriptors (Euclidean distance, as in the morphometric regressions above) of all runs present at each $t_D$, zero at the shared initial condition and rising as the organoids mature. The early bump reflects initial lumen formation, a large topological change to which SampEuler is highly sensitive. This change first happens in the high-$\xi$ runs, so the topology is briefly split. The slower runs then catch up as development progresses. The shaded band marks times at which fewer than $10$ runs remain, so fluctuations there reflect small-sample noise.}
\label{fig:divergence}
\end{figure}

All simulations start from the same initial conditions but develop distinct morphologies under different $(\tau_V,\xi)$ values (\Cref{fig:divergence}a). Panel (b) quantifies this divergence using the mean pairwise Euclidean distance between vectorised SampEuler descriptors at each model time $t_D$. The distance is zero initially and generally increases with development, consistent with improved parameter recovery at later stages. The early peak reflects differences in the timing of initial lumen formation: high-$\xi$ runs form lumina sooner, briefly increasing the spread, which then decreases as slower runs also form lumina. All $120$ runs contribute during this initial transient. In the shaded late-time region, fewer than ten runs remain, making the mean more sensitive to individual simulations.

\subsection*{Synthetic validation: 3-parameter ABC rejection}

\begin{figure}[htbp]
\centering
\includegraphics[width=\textwidth]{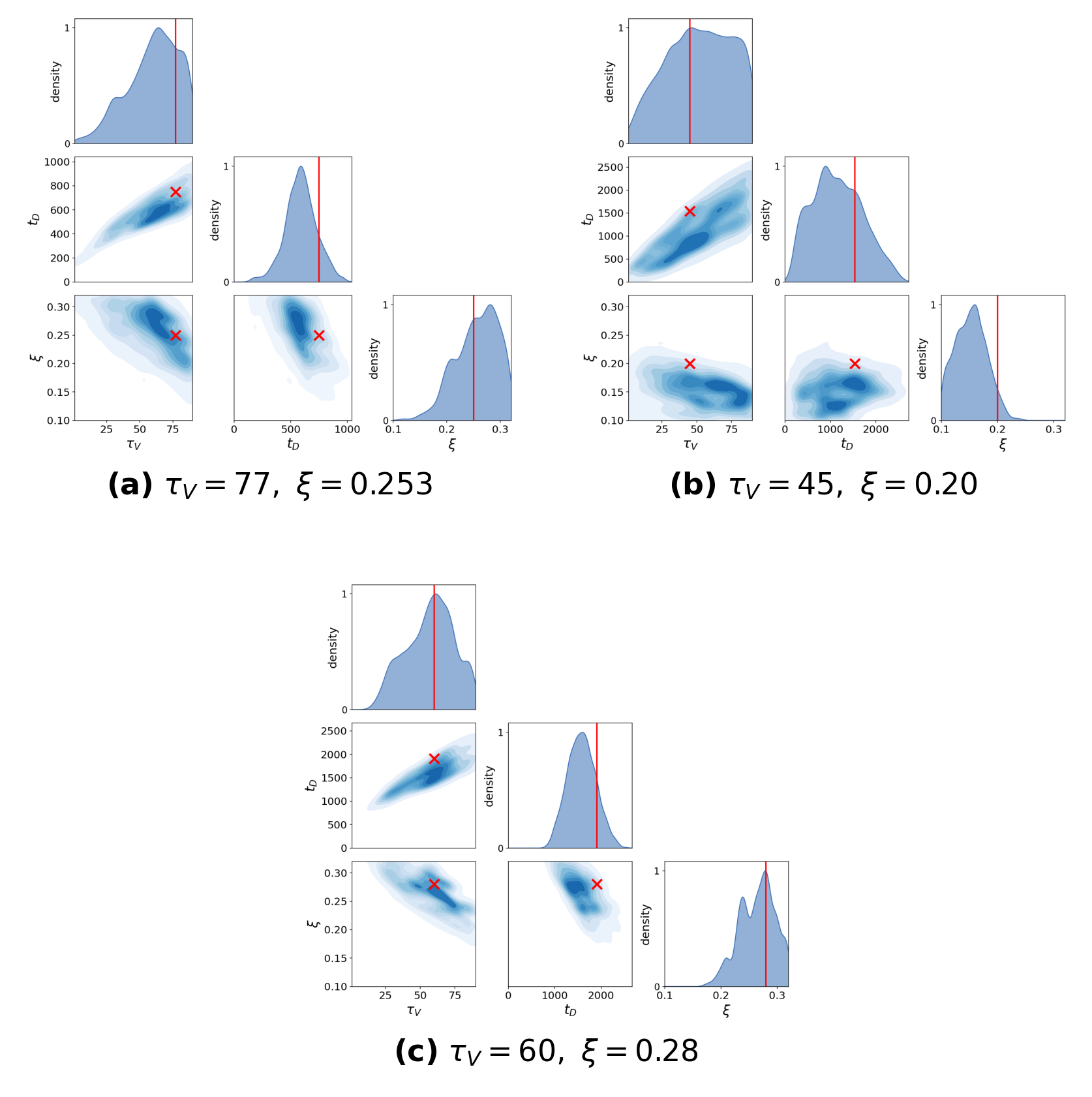}
\caption[Synthetic three-parameter ABC-rejection posteriors.]{\textbf{Synthetic 3-parameter ABC-rejection posteriors.} Joint $(\tau_V, \xi, t_D)$ rejection-ABC posteriors for the three synthetic organoids, each presented to the pipeline as a single end-time snapshot. Off-diagonal panels are pairwise posterior densities with the ground truth marked by a red cross. Diagonal panels are the normalised single-parameter marginal posteriors (red line marks the truth). The accepted set is the closest $1\%$ of the $60{,}000$ candidate frames under the SampEuler Wasserstein distance.}
\label{fig:synth_corners}
\end{figure}

To validate the inference pipeline, we apply it to three synthetic organoids generated from randomly drawn values of parameters $(\tau_V, \xi)$. To mimic the biological experiments, we take only the final frame of each run as the test input, and require that the pipeline recovers the developmental time $t_D$ at which the snapshot was taken together with the two model parameters. \Cref{fig:synth_corners} shows the resulting joint posteriors for$(\tau_V, \xi, t_D)$.

In all three cases the ground truth lies within the recovered posterior. The lumen osmotic pressure $\xi$ is the best-constrained parameter, with a unimodal one-dimensional marginal in every panel. The cell-cycle time $\tau_V$ and the developmental time $t_D$ have broad marginals, but their joint density is concentrated on a strongly positive correlated ridge, an elongated region of high posterior density, in the $(\tau_V, t_D)$ plane that passes through the ground truth, with $\xi$ decreasing as $\tau_V$ increases.

\begin{figure}[htbp]
\centering
\includegraphics[width=\textwidth]{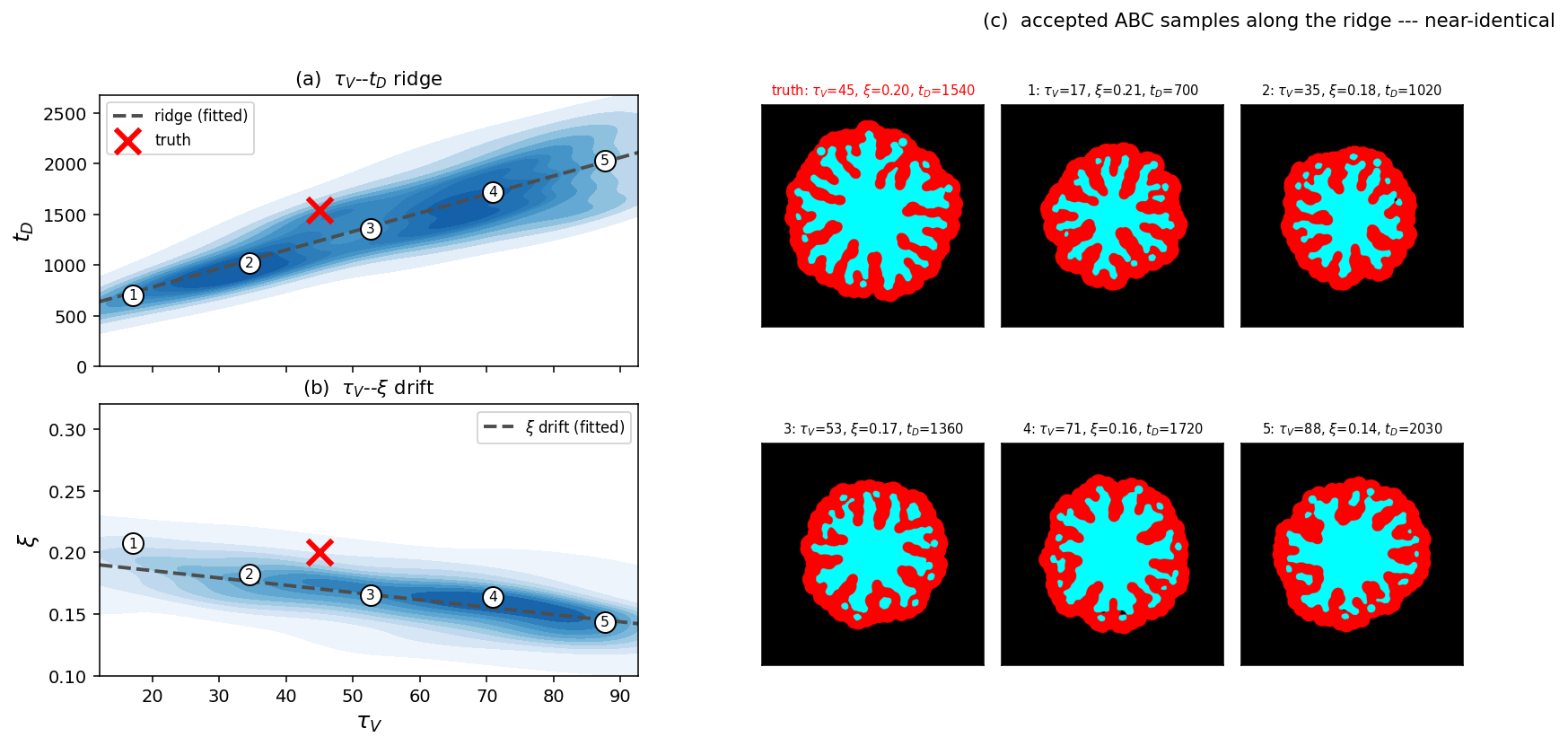}
\caption[The $\tau_V$--$t_D$ ridge.]{\textbf{Synthetic 3-parameter ridge.} The $\tau_V$--$t_D$ ridge for one organoid ($\tau_V = 45,\ \xi = 0.20$). \textbf{(a)}~The single-frame
posterior projected onto $(\tau_V, t_D)$, with a fitted line and five accepted posterior samples along it; the truth is the red cross. \textbf{(b)}~The same accepted samples in $(\tau_V, \xi)$: $\xi$ drifts downward as $\tau_V$ increases (fitted line), spanning $0.21$ to $0.14$ across the five samples. \textbf{(c)}~The observed (truth) organoid and the five accepted samples.}
\label{fig:ridge}
\end{figure}

We fit separate linear models to the accepted posterior samples to describe how $t_D$ and $\xi$ depend on $\tau_V$ along the ridge (\Cref{fig:ridge}a,b). We present the accepted simulations that are nearest to five evenly spaced points on the fitted line. Their organoids are morphologically near-indistinguishable from the observations (\Cref{fig:ridge}c). A single fixed-time image therefore constrains only the combination of cell-cycle time, osmotic pressure, and developmental time, but not each parameter separately: a faster-dividing organoid imaged earlier mimics a slower growing one imaged at later time. This trade-off explains the relation between the parameters. Combination of $(\tau_V, \xi, t_D)$ that lie on this one-dimensional line produce synthetic organoids with similar morphologies.

\subsection*{Temporal (trajectory) inference}
Although each experimental image represents a single timepoint, the simulator produces the full temporal trajectory associated with the earlier timepoints. For the synthetic organoids, we apply the same ABC-rejection pipeline to infer the model parameters from these trajectories. Because simulations contain different numbers of frames, we select 10 frames evenly spaced between the first and final timepoints of each simulation. Corresponding frames therefore represent the same fraction of each simulation’s duration. We compare these sequences using the temporal metric defined below. We accept the closest $5\%$ of trajectories here, compared with $1\%$ in the single-frame analysis. Each simulation contributes one candidate trajectory but many candidate frames, giving reference sets of $3{,}000$ trajectories and $60{,}000$ frames, respectively. We use a higher acceptance fraction for trajectories to retain more posterior samples from this smaller reference set.

\begin{definition}
  For two relative-time-aligned frame sequences $\{M_1, \dots, M_n\}$ and $\{N_1, \dots, N_n\}$ (here $n = 10$), the temporal SampEuler metric is
    $$d_{\mathrm{temp}} = \max_{i} d_{W}\big(\text{SampEuler}(M_i),\, \text{SampEuler}(N_i)\big),$$
  where $d_{W}$ is the Wasserstein distance between the SampEulers.
\end{definition}

\begin{figure}[htbp]
\centering
\includegraphics[width=\textwidth]{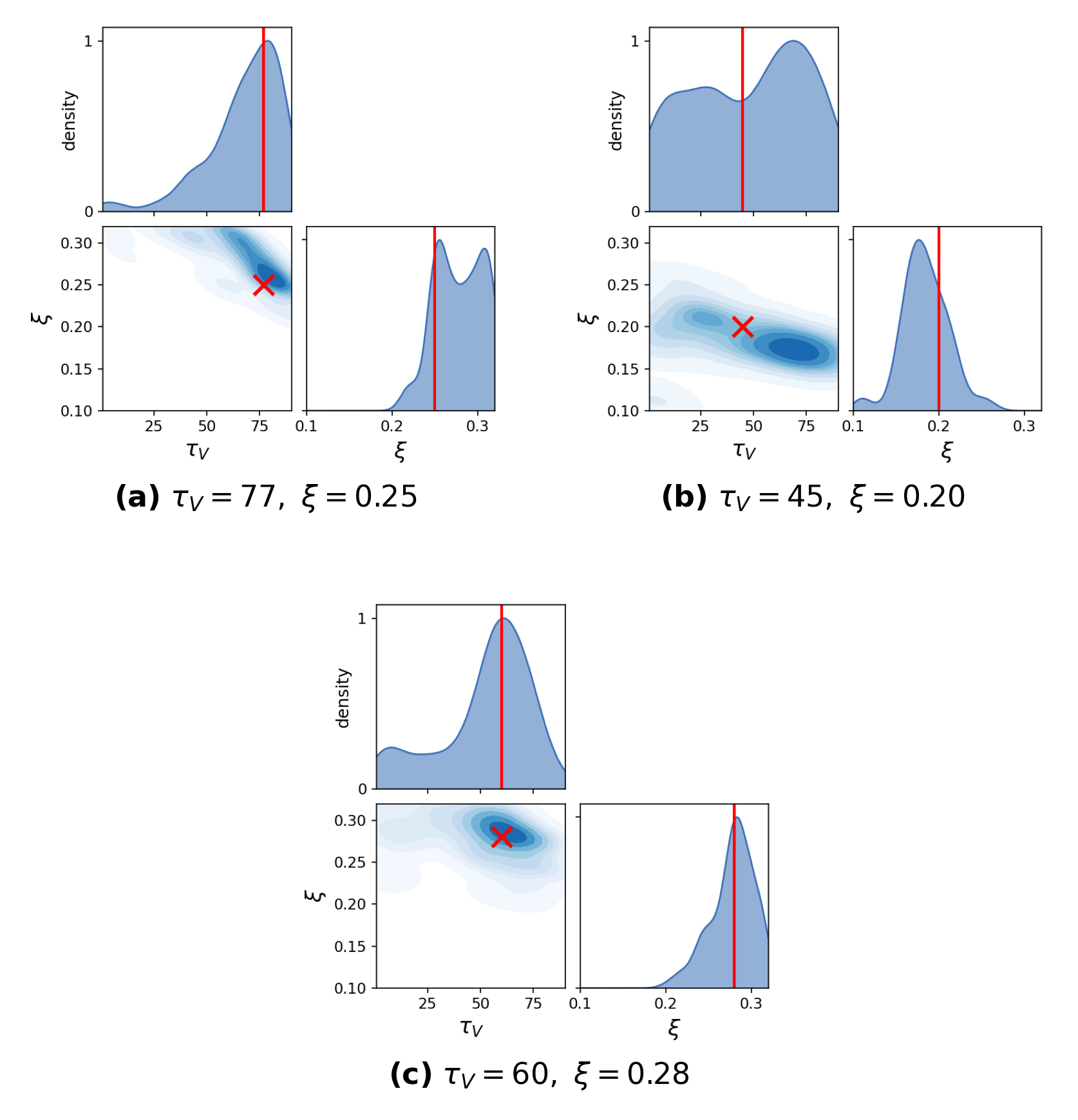}
\caption[Synthetic temporal-inference posteriors.]{\textbf{Synthetic 2-parameter posteriors.} Rejection-ABC posteriors over $(\tau_V, \xi)$ from temporal (10-frame) inference on the same three synthetic test organoids as \Cref{fig:synth_corners}, each shown as a corner plot: the joint $(\tau_V, \xi)$ density (filled contours, ground truth a red cross) with the normalised single-parameter marginal posteriors for each parameter on the diagonals (red line marks the truth). The accepted set is the closest $5\%$ of the $3{,}000$ simulated trajectories.}
\label{fig:temporal}
\end{figure}

\Cref{fig:temporal} shows the resulting $(\tau_V, \xi)$ posteriors under the temporal SampEuler metric. The ground truth lies within the recovered posterior for all three organoids. The ridge between $t_D$ and $\tau_V$ is removed in the temporal setting. What remains is the relationship between the two physical parameters: the temporal posteriors still carry a negative $(\tau_V, \xi)$ correlation, the analogue of the single-frame ridge in this $t_D$-free setting.

\begin{figure}[htbp]
\centering
\includegraphics[width=\textwidth]{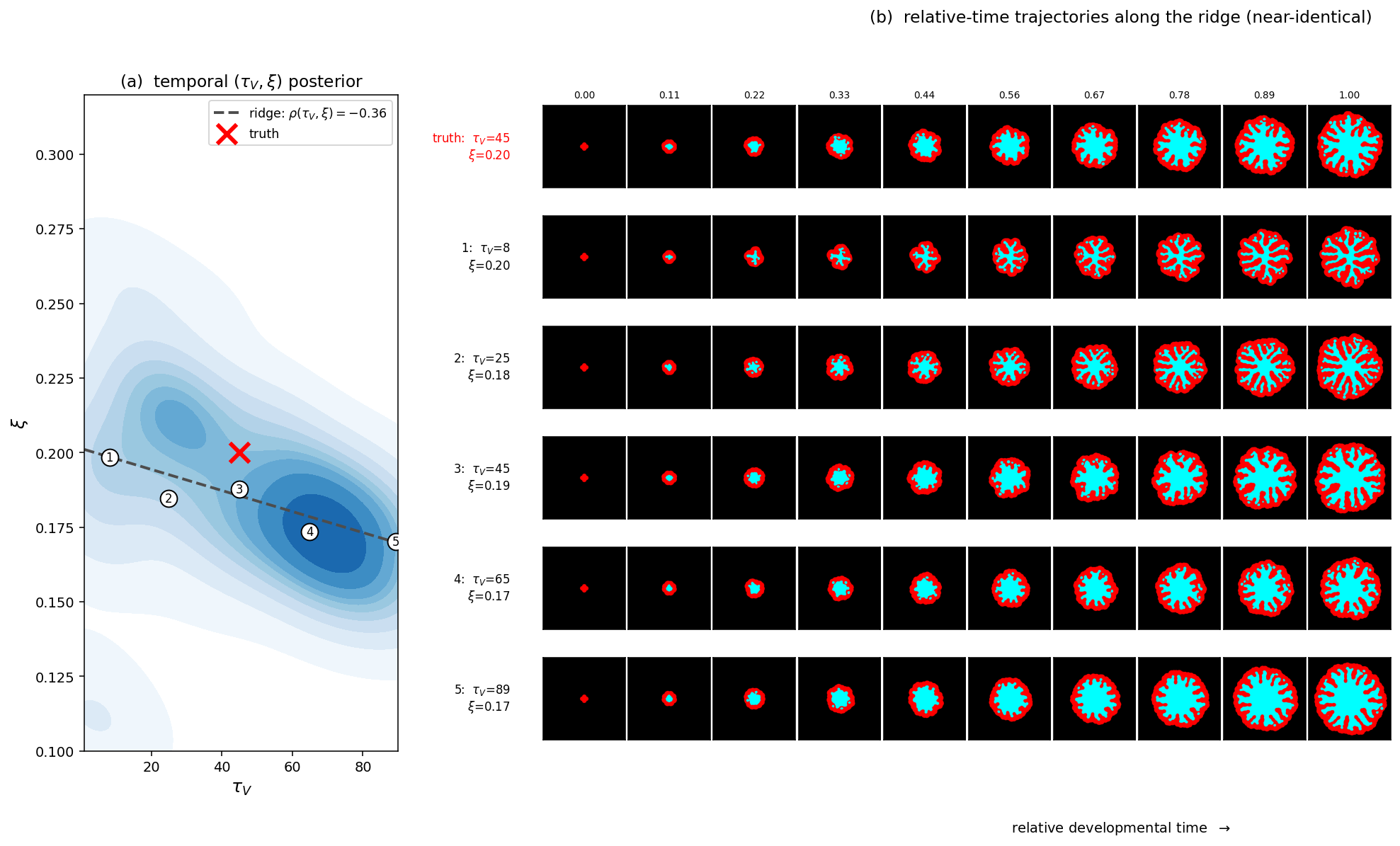}
\caption[The $\tau_V$--$\xi$ ridge under temporal inference.]{\textbf{Synthetic 2-parameter ridge.} The $(\tau_V, \xi)$ ridge under the trajectory inference, for the same organoid as \Cref{fig:ridge} ($\tau_V = 45,\ \xi = 0.20$). \textbf{(a)}~The temporal $(\tau_V, \xi)$ posterior with a fitted line (dashed), five accepted samples spaced along it, and the ground truth (red cross). \textbf{(b)}~The subsampled trajectories of those five samples (rows 2--6) and of the observed organoid (row 1).}
\label{fig:temporal_ridge}
\end{figure}

We examine the $(\tau_V, \xi)$ relation closely for the organoid used to generate the single-frame ridge in \Cref{fig:temporal_ridge}. For the accepted trajectories we fit a linear model to determine how $\xi$ varies with $\tau_V$ (\Cref{fig:temporal_ridge}a), and present the accepted simulations nearest to five data points, together with the observed organoid's trajectory (\Cref{fig:temporal_ridge}b). The six trajectories are near-indistinguishable even though $\tau_V$ spans $8$ to $89$. Different parameter pairs across this region of $(\tau_V, \xi)$ yield similar morphological trajectories. We conclude that the model preserves the robustness of the underlying biology, in which a range of similar $(\tau_V, \xi)$ combinations gives rise to similar organoid development. These findings are consistent with those of Lee et al.~\cite{lee2026permeability}, who found that lumen morphology is determined by the balance between cell proliferation and lumenal pressure. The ridge we identify is the set of $(\tau_V, \xi)$ combinations that preserve this balance.

\subsection*{Experimental inference: 3-parameter ABC rejection on pancreatic organoids}

\begin{figure}[htbp]
\centering
\includegraphics[width=\textwidth]{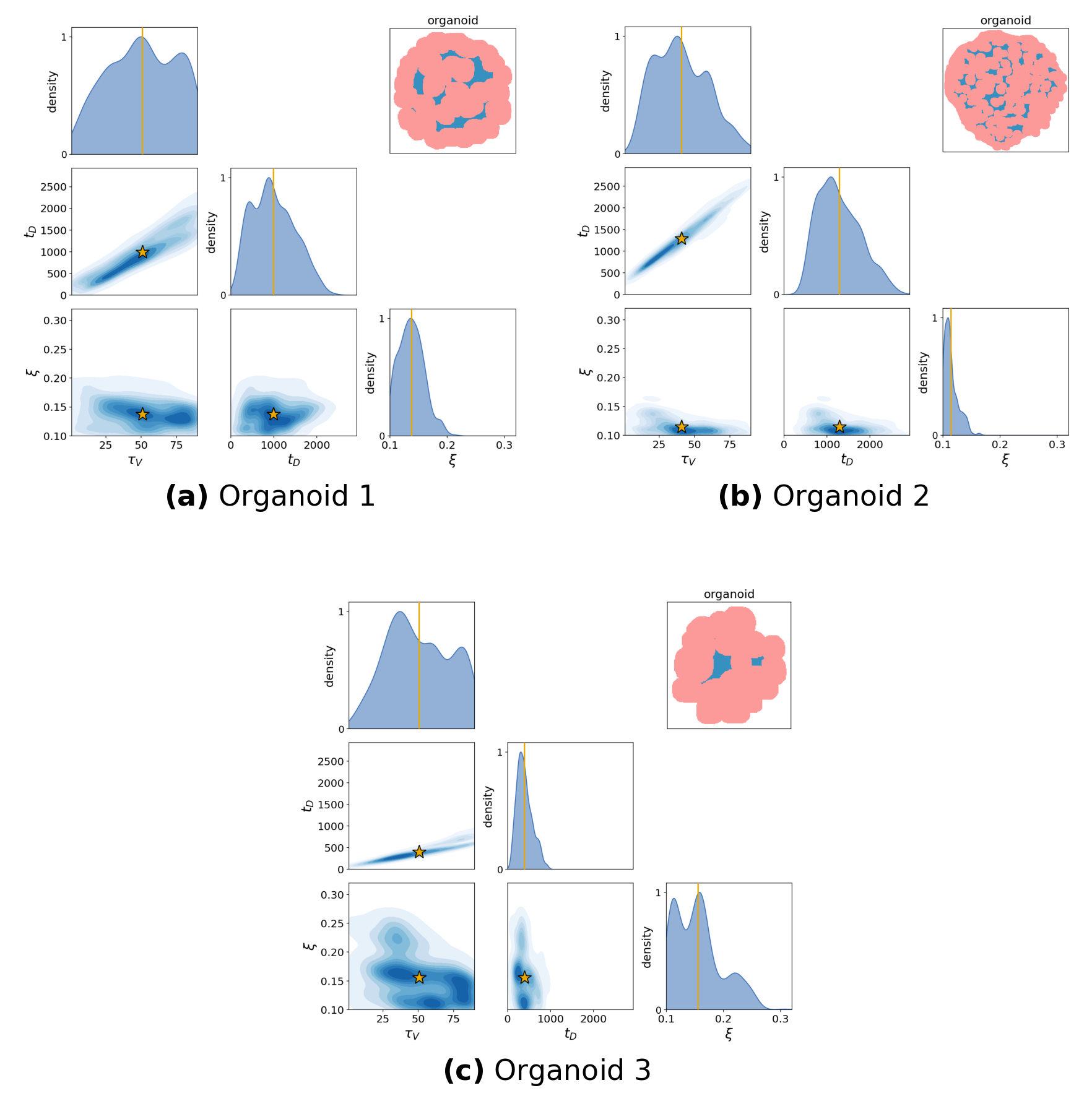}
\caption[Experimental three-parameter ABC-rejection posteriors.]{\textbf{Experimental 3-parameter ABC-rejection posteriors.} Joint $(\tau_V, \xi, t_D)$ ABC-rejection posteriors for three representative pancreatic organoids, each presented as a single fixed-time snapshot. Off-diagonal panels are pairwise posterior densities with the star marking the posterior mean (a point estimate using the posterior distribution, as the experimental images have no ground truth). Diagonal panels are the normalised single-parameter marginal posteriors (gold line at the posterior mean). The corresponding organoid images are shown at the top right (red is cells, blue is lumen). The accepted set is the closest $1\%$ of the $60{,}000$ candidate frames under the SampEuler Wasserstein distance.}
\label{fig:exp_corners}
\end{figure}

We now apply the pipeline to 10 images from pancreatic organoids, each a two-dimensional cross-section in the plane of largest area. Three illustrative examples of distinct morphologies are presented in \Cref{fig:exp_corners}, and the remaining seven organoids are presented in \nameref{S1_Fig}.

Organoid 1 in \Cref{fig:exp_corners}(a) and the seven in \nameref{S1_Fig} share similar lumen geometries, and the pipeline infers similar $(\tau_V, \xi)$ posteriors for them, confirming the inference is consistent across the group. Organoid 2 in \Cref{fig:exp_corners}(b) and Organoid 3 in \Cref{fig:exp_corners}(c) are distinct from this common type, and are discussed further below. 

Organoid 2 in \Cref{fig:exp_corners}(b) forms a complex, branching lumen, and its posterior favours a very low osmotic pressure $\xi$. This is consistent with the experimental finding of \cite{lee2026permeability} that reducing the lumen pressure transforms spherical lumina into complex, branching ones. 

The posterior distribution for Organoid 3 in \Cref{fig:exp_corners}(c) place it at an early developmental time $t_D$, consistent with its simple lumen topology and its small number of cells. All ten experimental organoids were imaged at the same experimental age. This organoid was initialised with a smaller number of cells which proliferated more slowly than the other organoids. Consequently, it is the least developed of the group when imaged. Its broad $(\tau_V, \xi)$ posterior reflects that early developmental stages give near-identical morphology regardless of the conditions $\tau_V$ and $\xi$.

\section*{Discussion}
We have introduced a pipeline that infers the principal physical drivers of pancreatic organoid morphology from image data. The framework combines topological data analysis with likelihood-free inference, linking experimentally observed shape to the parameters of a mechanistic model. Each organoid is summarised by SampEuler, a shape descriptor derived from the Euler characteristic transform and adapted to organoid images through a weighted cubical complex. We first demonstrated, using synthetic organoids generated by a phase-field model, that SampEuler captures relevant morphological information. For synthetic organoids, linear regression models recover conventional morphological descriptors used in previous studies. 
Nearest-neighbour matching using the Wasserstein distance between SampEuler summaries also retrieved simulations with parameter values close to those used to generate the synthetic test images. This demonstrates that the descriptor contains information relevant to parameter inference. Building on the relationships between morphology and model parameters established by Lee et al., our ABC-rejection framework estimates approximate joint posterior distributions over cell-cycle time $\tau_V$, lumen osmotic pressure $\xi$, and developmental time $t_D$ for individual organoids. In synthetic tests, the inferred posteriors include the true parameter combinations. For experimental organoids, they identify parameter regimes compatible with the observed morphologies under the model.

A central finding is that the inverse problem reveals which combinations of the three parameters give morphologically equivalent output. Inference from a single snapshot therefore returns a ridge-shaped posterior in $(\tau_V, \xi, t_D)$ space rather than a unique estimate. One reason for this is as follows: a single image contains no direct information about developmental age, so an organoid with a short cell-cycle time observed early in development may resemble one with a longer cell-cycle time observed at a later timepoint. A second source of non-identifiability becomes apparent only when temporal information is incorporated. When we infer parameters from complete developmental trajectories, developmental time is no longer a free parameter, reducing the inference problem to $(\tau_V,\xi)$. The resulting posteriors reveal a negative relationship between the cell-cycle time and osmotic pressure: an increase in the value of $\tau_V$ can be compensated by a reduction in the value of $\xi$ while preserving overall morphology. This relationship appears to reflect a genuine property of the underlying biology rather than an artefact of the inference procedure. In particular, it supports and extends the conclusions of Lee et al., who showed that lumen morphology is determined primarily by the balance between proliferation and osmotic pressure~\cite{lee2026permeability}.

Applied to experimental organoids, the pipeline yields posteriors that are consistent with qualitative biological expectations. Most organoids have similar lumen architectures and are assigned overlapping regions of parameter space, suggesting that inference is reproducible across morphologically similar samples. By contrast, two morphologically distinct organoids occupy different regions of parameter space. The branching organoid (\Cref{fig:exp_corners}b) is associated with relatively low inferred osmotic pressure, in agreement with the finding of Lee et al.\ that branching organoids have lower lumen pressure than spherical organoids~\cite{lee2026permeability}. Similarly, the least developed organoid (\Cref{fig:exp_corners}c) is assigned a posterior concentrated at earlier developmental times, consistent with its immature morphology. Together, these results indicate that the inferred parameter values are consistent with independent qualitative observations of the organoids.

Two limitations of our study arise from the phase field model. The first concerns model fidelity. The phase-field framework of~\cite{tanida2025predicting,lee2026permeability} was developed to reproduce organoid morphology and necessarily incorporates simplifying assumptions. For example, regularisation terms favour smooth lumen boundaries and relatively simple cell shapes. In practice, experimental organoids are more irregular: cells can be elongated and may interlock, and lumina often contain sharp corners and fine structural detail. Consequently, some discrepancies between simulated and observed morphology are inevitable. A second issue concerns the initial conditions. All simulations begin from the same small cellular aggregate, and the initial state is assumed to be independent of the cell cycle time and osmotic pressure. Experimental organoids are likely to originate from different initial cell numbers and configurations. A smaller founding aggregate, for example, may appear less developed at a given age simply because fewer rounds of proliferation have occurred (see Organoid 3 in \Cref{fig:exp_corners}c). Since the pipeline normalises overall size and assumes the same initial conditions, part of the inferred variation in $(\tau_V,\xi)$ may reflect differences in initial conditions rather than genuine biological differences in proliferation or osmotic pressure. Extending the mechanistic model to include variable initial conditions therefore represents a natural next step.

The second limitation concerns computational expense. We adopted ABC rejection because it allows a single simulation library to be reused for multiple observations. More sophisticated likelihood-free methods, such as ABC-SMC, were not considered because of the cost of generating additional simulations. A faster simulator would make such approaches feasible. Importantly, however, these limitations arise from the phase field model rather than the inference framework itself. The topology-based inference pipeline is agnostic to the underlying simulator and could be coupled to alternative mechanistic models as they become available. One natural alternative is an agent-based representation~\cite{osborne2017comparing, van2015simulating, graner1992simulation}, which resolves individual cells. Such models may capture irregular cell and lumen geometries more faithfully and, depending on implementation, may also reduce computational cost. Established platforms such as Chaste~\cite{cooper2020chaste, kursawe2018approximate} would facilitate their development.

Several further model extensions are possible. The most obvious is to move from two-dimensional to three-dimensional data. Three-dimensional phase-field simulations have already been demonstrated~\cite{tanida2025predicting}, and volumetric imaging of organoids is feasible~\cite{lee2026permeability}. Since the Euler characteristic transform is defined in arbitrary dimension, SampEuler extends naturally to three-dimensional morphologies. The principal obstacle is computational rather than methodological: three-dimensional simulations are substantially more expensive than their two-dimensional counterparts. A second extension is the analysis of experimental time-series data. As our trajectory-based experiments demonstrate, temporal information substantially improves parameter identifiability and would sharpen inference of $(\tau_V,\xi)$.

Beyond these methodological extensions, the framework could be used to infer other biological parameters directly from morphology, provided they are encoded in the model. These might include the effect of a drug treatment, a genetic perturbation, or the mechanical properties of the extracellular matrix. Inferring such parameters would complement existing approaches for comparing organoid responses to genetic and chemical perturbations~\cite{zhou2026development,zhou2026identifying}.

In summary, we have shown that key mechanistic drivers of organoid morphology can be inferred from shape alone. By combining topological summaries with likelihood-free inference, we recover information about proliferation, osmotic pressure, and developmental stage directly from organoid images. More broadly, the work illustrates how topology can serve as an information-preserving bridge between biological morphology and mechanistic modelling. Although demonstrated here for pancreatic organoids, the approach is not specific to this system and may be applicable to a wide range of organoids and developing tissues in which morphology reflects underlying physical and biological processes.

\clearpage   
\section*{Supporting information}


\paragraph*{S1 Appendix.}\label{S1_Appendix}
{\bf Mathematical Definition of SampEuler.}

We first define geometric simplicial complexes, the mathematical framework for the shapes with which we work. We define the dimension of a simplex and then use it to define the Euler characeristic, the Euler characteristic transform (ECT) and the related SampEuler.

\begin{definition}
    A \textbf{geometric $k$-simplex} $K_1$ is the convex hull of $k+1$ affinely independent points $v_0,v_1,\dots, v_k$ and denoted by $[v_0,\dots, v_k]$. For any $\{u_0,\dots,u_j\}\subset\{v_0,\dots,v_k\}$, we call $[u_0,\dots,u_j]$ a \textbf{face} of $[v_0,v_1,\dots, v_k]$. The \textbf{dimension} of $K_1$ is $\dim(K_1) = k$.

    A geometric simplicial complex $K$ is a finite collection of geometric simplices $\{K_1,\dots,K_n\}$ such that
    \begin{enumerate}
        \item For every $K_i \in K$, every face of $K_i$ is also an element of $K$;
        \item For any $K_i, K_j \in K$, either $K_i \cap K_j = \emptyset$ or $K_i \cap K_j$ is a face of both $K_i$ and $K_j$.
    \end{enumerate}
    The \textbf{dimension} of $K$ is defined as $\max\{\dim(K_i):i\in\{1\dots n\}\}$.
\end{definition}

\begin{definition}
    For a geometric $k$-simplex $K_1$, let $K_1^{\circ}$ denote its relative interior. The Euler characteristic of its relative interior is defined to be $\chi(K_1^{\circ}):=(-1)^k$. For a geometric simplicial complex $K$ consisting of the finite collection of simplices $\{K_1,\dots,K_n\}$ with disjoint relative interiors, the Euler characteristic of $K$ is defined as $\chi(K) = \sum_{i=1}^n \chi(K_i^{\circ})$.
\end{definition}

\begin{definition}
    For a geometric simplicial complex $K\subset\R^d$ consisting of simplices $K_1,\dots, K_n$, the \textbf{Euler characteristic transform} of $K$, denoted by $\ect(K): S^{d-1}\times\R\rightarrow\Z$, is defined as:
        $$\ect(K)(\mathbf{v},t) = \chi(K^{\langle \mathbf{v}, \cdot\rangle\leq t}),$$
    where $K^{\langle \mathbf{v}, \cdot\rangle\leq t} = \{K_i: \langle \mathbf{v}, x\rangle\leq t, \forall x\in K_i\}$ denoting the set of simplices in $K$ that fully lies below the hyperplane defined by $\mathbf{v}$ and $t$. Fixing direction $\mathbf{v}$, the curve $\ect(K)(\mathbf{v},-):\R\rightarrow\Z$ is the \textbf{Euler characteristic curve (ECC)} of $K$ in the direction $\mathbf{v}$.
\end{definition}

Following Kendall~\cite{kendall1999shape}, we take the shape of a complex to be its equivalence class under rotations, reflections, and translations. Distinct complexes have distinct ECTs~\cite{turner2014persistent, curry2022many}. However, rotated or reflected copy of a complex can have a different ECT while having the same shape. Centring removes translation, but there is no canonical way to remove rotations and reflections. We therefore consider the ECT pushforward measure. It records the distribution of Euler characteristic curves obtained as the direction varies over the whole unit sphere. The ECT pushforward measure is invariant under rotation and reflection. To make this precise, the ECT of a geometric simplicial complex $K$ can be viewed as a function from the unit sphere $S^{d-1}$ to the space of integer-valued functions on $\R$ with finitely many jumps, denoted by $\CF$. One can show that this function is continuous with respect to the $L^1$-distance on $\CF$ and the Euclidean distance on $S^{d-1}$~\cite{curry2022many}. Continuity is what makes the pushforward of the uniform measure $\mu$ on $S^{d-1}$ well defined.

\begin{definition}
    For geometric simplicial complexes $K\subset\R^d$, the \deff{ECT pushforward measure} of $K$ is $\ect(K)_*\mu: \mathcal{B}(\CF)\rightarrow \R$, such that $\ect(K)_*\mu(E) = \mu(\{v\in S^{d-1}: \ect(K)(v,-)\in E\})$, where $\mu$ is the uniform measure on the sphere and $\mathcal{B}(\CF)$ is the Borel $\sigma$-algebra on the space of ECCs.
\end{definition}

\begin{definition}
Let $K\subset\R^d$ be a geometric simplicial complex and let $\ect(K)_*\mu$ be the ECT pushforward measure of $K$.  Randomly sample $(X_1,\dots,X_n)\stackrel{\mathrm{i.i.d.}}{\sim} \ect(K)_*\mu.$
The \emph{SampEuler} of $K$ of order $n$, denoted $\samp_n$, is defined by
$$
  \samp_n(K) \;=\; \frac{1}{n}\sum_{i=1}^n \delta_{X_i},
$$
where $\delta_x$ is the Dirac measure at $x$.
\end{definition}

For generic geometric simplicial complexes, the ECT pushforward measure acts as a complete shape descriptor and provides an invertible signature~\cite{curry2022many, ghrist2018persistent}. Further, SampEuler converges to the ECT pushforward measure as $n\rightarrow\infty$~\cite{yang2026topological}. This justifies the use of SampEuler as a shape descriptor for complex organoid morphologies.

 We next describe how to construct a finite-dimensional feature vector from SampEuler for use in statistical models.

\begin{definition}
Let $K$ be an embedded simplicial complex in $\R^d$ and $l >0$. Denote by $\Int(\R)_l$ the set of intervals of length $l$ and $K_{[a,b),c}: = \{\phi\in \text{supp}(\samp_n(K)): \phi|_{[a,b)} = c\1_{[a,b)}\}$ for $a,b\in\R$ and $c\in\Z$. We define the \deff{vectorization of SampEuler}(or vectorized SampEuler) of $K$ as follows:
    \begin{align*}
        \mathcal{V}(K, l): \Int(\R)_l\times\Z &\rightarrow [0,1],\\
        ([a,b),k) &\mapsto \samp_n(K)(K_{[a,b),k}),
    \end{align*}
    The parameter $l$ is the bin width. It sets the resolution at which the filtration axis is discretised. We abbreviate the notation to $\mathcal{V}(K)([a,b),k)$ when $l$ is clear from the context. 
\end{definition}

The entry $\mathcal V(K,l)(I,k)$ records the proportion of sampled curves that remain constant at Euler characteristic $k$ over the interval $I$. To obtain a finite representation, we partition a bounded filtration range into $m$ intervals of length $l$ and select integer values $k_{\min},\dots, k_{\max}$. Evaluating $\mathcal V(K,l)$ at each interval–value pair gives an $(m\times(k_{\max}-k_{\min}+1))$-dimensional matrix. Flattening this matrix in a fixed order yields a feature vector of length $m(k_{\max}-k_{\min}+1)$.

The above definitions are stated for simplicial complexes. A cubical complex is constructed in the same way, but from squares glued along shared edges and vertices rather than triangles. The definitions for simplicial complexes generalise naturally for cubical complexes. For weighted complexes, we define the \textbf{weighted Euler characteristic} of a square, edge or vertex $s$ as $w(s)\cdot\chi(s^{\circ})$, and the sum over all of the simplices is the weighted Euler characteristic of $K$. The ECT of the weighted cubical complex $K$ are then defined as follows:
$$
\ect(K)(v,t) \;=\; \sum_{\sigma\in K^{\langle v,\cdot\rangle\le t}} w(\sigma)\,\chi(\sigma^{\circ}).
$$
Similarly, SampEuler is defined as the empirical measure of the weighted Euler characteristic curves. An efficient implementation for integer-valued images is provided by~\cite{lebovici2024efficient}. We choose a bounded filtration interval for all ECCs. We use the $1$-Wasserstein distance with the $L^1$ ground metric on $\CF$ to compare SampEuler empirical measures. \Cref{ex:sampeuler} illustrates the pipeline on a worked example.

\newpage

\paragraph*{S1 Fig.}\label{S1_Fig}
{\bf Experimental ABC-rejection posteriors for the remaining seven organoids.}
Joint $(\tau_V, \xi, t_D)$ ABC-rejection posteriors for the seven pancreatic
organoids not shown in \Cref{fig:exp_corners}. Plots and markers are as in
\Cref{fig:exp_corners}.

\begin{figure}[htbp]
\centering
\includegraphics[width=\textwidth]{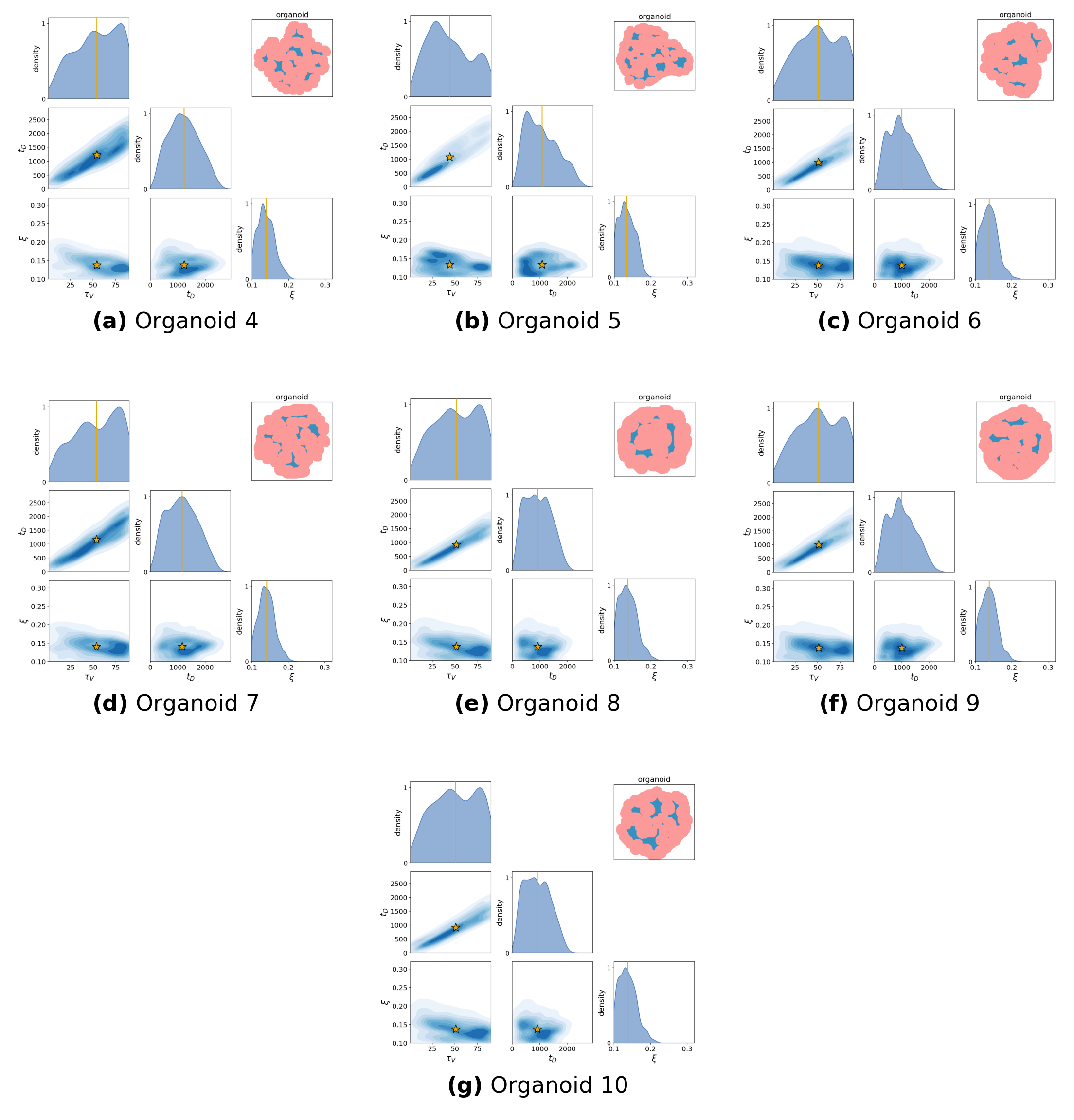}
\end{figure}

\newpage

\paragraph*{S2 Appendix.}
\label{S2_Appendix}
{\bf Phase-field organoid model.}
Here we summarise the forward simulator of~\cite{tanida2025predicting, lee2026permeability}, which those papers develop in full. Each cell $u_i$ ($i=1,\dots,M$) and the lumen $u_0$ are smooth fields taking values in $[0,1]$ on the domain $\Omega$, and all fields evolve by gradient descent on a shared free energy, $\tau\,\partial_t u_i = -\,\delta F/\delta u_i$, where $\tau$ is the field relaxation time (distinct from the cell-cycle time $\tau_V$ below) and
\begin{equation}\label{eq:si-free-energy}
\begin{aligned}
F = {} & \sum_{i=0}^{M} \int_{\Omega}\!\Big[\tfrac{D}{2}|\nabla u_i|^2 + \tfrac{1}{4}u_i^2(1-u_i)^2\Big]\,d\vec r
 + \sum_{i\neq j}\tfrac{\beta}{12}\!\int_{\Omega}\! h(u_i)\,h(u_j)\,d\vec r
 + \sum_{k=1}^{M} \tfrac{\alpha}{12}\big(V^{\mathrm{max}}_k - V_k\big)^2 \\
 & + \sum_{k\neq l}\tfrac{\eta}{12}\!\int_{\Omega}\!\nabla h(u_k)\!\cdot\!\nabla h(u_l)\,d\vec r
 + \sum_{k=1}^{M} \tfrac{\gamma}{12}\!\int_{\Omega}\!|\nabla h(u_k)|^2\,d\vec r
 - \tfrac{\xi}{6}\!\int_{\Omega}\! h(u_0)\,d\vec r ,
\end{aligned}
\end{equation}
where $h(u)=u^2(3-2u)$ is a smooth indicator, $V_k=\int_\Omega h(u_k)\,d\vec r$ is the volume of cell $k$, $V^{\mathrm{max}}_k$ is its (time-dependent) target volume (\eqref{eq:si-tauv}), and the coefficients $D,\alpha,\beta,\eta,\gamma$ set, in order, the interface width, cell-volume control, volume exclusion, cell--cell adhesion, and boundary regularity.

The two inferred parameters enter as follows. The osmotic pressure $\xi$ is the coefficient of the lumen term in \eqref{eq:si-free-energy}: this term lowers $F$ as the lumen expands, so a larger $\xi$ more strongly favours lumen inflation. The cell-cycle time $\tau_V$ sets the pace of growth: the $\alpha$-term drives each cell volume $V_k$ towards a per-cell target $V^{\mathrm{max}}_k$, which itself grows towards the fixed value $\bar{V}$ with timescale $\tau_V$,
\begin{equation}\label{eq:si-tauv}
\tau_V\,\frac{d V^{\mathrm{max}}_k}{dt} = \bar{V} - V^{\mathrm{max}}_k ,
\end{equation}
so a smaller $\tau_V$ grows cells faster and shortens the interval between divisions, whereas a larger $\tau_V$ slows growth~\cite{tanida2025predicting, lee2026permeability}. A cell divides once its volume passes the threshold $\bar{V}-v_d$. It splits along a division plane set by a spindle-pole model and seeds a new microlumen~\cite{tanida2025predicting}. Each cell is assigned its own growth timescale $\tau_{V,k} = \tau_V(1+\zeta)$ when it is created, with $\zeta$ drawn uniformly from $[-0.25, 0.25]$, so cells reach the division threshold at different times. The inferred $\tau_V$ is a population mean rather than a per-cell constant. Each run integrates forward from four cells until either a hard step cap is reached or an early stop is triggered when the lumen leaks through the cell layer. Both the growth and the leakage time depend on $(\tau_V,\xi)$, and so does the developmental time a run reaches. All remaining coefficients are held at the defaults of~\cite{tanida2025predicting}: $D=0.001$, $\alpha=1$, $\beta=1$, $\gamma=0.01$, $\eta=0.008$, $\bar{V}=3.0$, and $v_d=0.1$, giving a division threshold of $2.9$.





\section*{Data and code availability}
All analysis code and the experimental organoid images used in this study are available at \url{https://github.com/reddevil0623/organoid_inference}. The phase-field forward simulator is the published C\texttt{++} model of~\cite{tanida2025predicting, lee2026permeability}. The ABC reference simulation pool is not archived here owing to its size, but is fully regenerable from the provided scripts.

\section*{Acknowledgments}
HAH and HMB are grateful for the support provided by the UK Centre for Topological Data Analysis Engineering and Physical Sciences Research Council (EPSRC) grant EP/R018472/1 and EPSRC EP/Z531224/1. HAH gratefully acknowledges funding from the Royal Society RGF/EA/201074, UF150238 and EPSRC EP/Y028872/1 and EP/Z531224/1. HAH and HY acknowledge funding by the Leverhulme Trust Prize PLP-2020-252. HY is grateful for the support of the Leathersellers' Foundation. HMB acknowledges support provided by the Mark Foundation for Cancer Research. The biological data and phase field model were generated with funding of the Human Frontiers Science Program project number RGP0050\slash2018.

\nolinenumbers

%
%
%
\bibliography{bibliography}

@article{tanida2025predicting,
  title={Predicting organoid morphology through a phase field model: insights into cell division and lumenal pressure},
  author={Tanida, Sakurako and Fuji, Kana and Lu, Linjie and Guyomar, Tristan and Lee, Byung Ho and Honigmann, Alf and Grapin-Botton, Anne and Riveline, Daniel and Hiraiwa, Tetsuya and Nonomura, Makiko and others},
  journal={PLOS Computational Biology},
  volume={21},
  number={8},
  pages={e1012090},
  year={2025},
  publisher={Public Library of Science San Francisco, CA USA}
}

@article{lee2026permeability,
  title={Permeability-driven pressure and cell proliferation control lumen morphogenesis in pancreatic organoids},
  author={Lee, Byung Ho and Fuji, Kana and Petzold, Heike and Seymour, Phil and Yennek, Siham and Schewin, Coline and Lewis, Allison and Riveline, Daniel and Hiraiwa, Tetsuya and Sano, Masaki and Grapin-Botton, Anne},
  journal={Nature Cell Biology},
  volume={28},
  number={1},
  pages={113--124},
  year={2026},
  publisher={Nature Publishing Group}
}

@article{toni2009,
  title={Approximate {B}ayesian computation scheme for parameter inference and model selection in dynamical systems},
  author={Toni, Tina and Welch, David and Strelkowa, Natalja and Ipsen, Andreas and Stumpf, Michael P H},
  journal={Journal of the Royal Society Interface},
  volume={6},
  number={31},
  pages={187--202},
  year={2009},
  publisher={The Royal Society}
}

@article{delmoral2012abcsmc,
  title={An adaptive sequential {M}onte {C}arlo method for approximate {B}ayesian computation},
  author={Del Moral, Pierre and Doucet, Arnaud and Jasra, Ajay},
  journal={Statistics and Computing},
  volume={22},
  number={5},
  pages={1009--1020},
  year={2012},
  publisher={Springer}
}

@article{beaumont2002abc,
  title={Approximate {B}ayesian computation in population genetics},
  author={Beaumont, Mark A and Zhang, Wenyang and Balding, David J},
  journal={Genetics},
  volume={162},
  number={4},
  pages={2025--2035},
  year={2002},
  publisher={Oxford University Press}
}

@article{beaumont2009adaptive,
  title={Adaptive approximate {B}ayesian computation},
  author={Beaumont, Mark A and Cornuet, Jean-Marie and Marin, Jean-Michel and Robert, Christian P},
  journal={Biometrika},
  volume={96},
  number={4},
  pages={983--990},
  year={2009},
  publisher={Oxford University Press}
}

@article{pritchard1999,
  title={Population growth of human {Y} chromosomes: a study of {Y} chromosome microsatellites},
  author={Pritchard, Jonathan K and Seielstad, Mark T and Perez-Lezaun, Anna and Feldman, Marcus W},
  journal={Molecular Biology and Evolution},
  volume={16},
  number={12},
  pages={1791--1798},
  year={1999},
  publisher={Oxford University Press}
}

@article{thorne2021tabc,
  title={Topological approximate {B}ayesian computation for parameter inference of an angiogenesis model},
  author={Thorne, Thomas and Kirk, Paul D W and Harrington, Heather A},
  journal={Bioinformatics},
  volume={38},
  number={9},
  pages={2487--2493},
  year={2022},
  publisher={Oxford University Press},
  note={Preprint at arXiv:2108.11640}
}

@article{buzbas2015aabc,
  title={{AABC}: approximate approximate {B}ayesian computation for inference in population-genetic models},
  author={Buzbas, Erkan O and Rosenberg, Noah A},
  journal={Theoretical Population Biology},
  volume={99},
  pages={31--41},
  year={2015},
  publisher={Elsevier}
}

@article{gutmann2016bolfi,
  title={{B}ayesian optimization for likelihood-free inference of simulator-based statistical models},
  author={Gutmann, Michael U and Corander, Jukka},
  journal={Journal of Machine Learning Research},
  volume={17},
  number={125},
  pages={1--47},
  year={2016}
}

@article{jarvenpaa2021parallel,
  title={Parallel {G}aussian process surrogate {B}ayesian inference with noisy likelihood evaluations},
  author={J{\"a}rvenp{\"a}{\"a}, Marko and Gutmann, Michael U and Vehtari, Aki and Marttinen, Pekka},
  journal={Bayesian Analysis},
  volume={16},
  number={1},
  pages={147--178},
  year={2021},
  publisher={International Society for Bayesian Analysis}
}

@misc{yang2026topological,
  title={Topological shape transform for thymus structures},
  author={Yang, Haochen and Lebovici, Vadim and Tarcevski, Andreas and Tchernev, Liliana and Zuklys, Saulius and Holl{\"a}nder, Georg A and Byrne, Helen M and Harrington, Heather A},
  year={2026},
  howpublished={arXiv:2602.18889}
}

@article{curry2022many,
  title={How many directions determine a shape and other sufficiency results for two topological transforms},
  author={Curry, Justin and Mukherjee, Sayan and Turner, Katharine},
  journal={Transactions of the American Mathematical Society, Series B},
  volume={9},
  number={32},
  pages={1006--1043},
  year={2022}
}

@InProceedings{lebovici2024efficient,
  author =	{Lebovici, Vadim and Oudot, Steve and Passe, Hugo},
  title =	{{Efficient Computation of Topological Integral Transforms}},
  booktitle =	{22nd International Symposium on Experimental Algorithms (SEA 2024)},
  pages =	{22:1--22:19},
  series =	{Leibniz International Proceedings in Informatics (LIPIcs)},
  ISBN =	{978-3-95977-325-6},
  ISSN =	{1868-8969},
  year =	{2024},
  volume =	{301},
  editor =	{Liberti, Leo},
  publisher =	{Schloss Dagstuhl -- Leibniz-Zentrum f{\"u}r Informatik},
  address =	{Dagstuhl, Germany},
  URL =		{https://drops.dagstuhl.de/entities/document/10.4230/LIPIcs.SEA.2024.22},
  URN =		{urn:nbn:de:0030-drops-203878},
  doi =		{10.4230/LIPIcs.SEA.2024.22}
}

@article{ghrist2018persistent,
  title={Persistent homology and Euler integral transforms},
  author={Ghrist, Robert and Levanger, Rachel and Mai, Huy},
  journal={Journal of Applied and Computational Topology},
  volume={2},
  pages={55--60},
  year={2018},
  publisher={Springer}
}

@book{robert2004monte,
  title={Monte Carlo statistical methods},
  author={Robert, Christian P and Casella, George},
  volume={2},
  year={2004},
  publisher={Springer}
}

@article{lancaster2014organogenesis,
  title={Organogenesis in a dish: modeling development and disease using organoid technologies},
  author={Lancaster, Madeline A and Knoblich, Juergen A},
  journal={Science},
  volume={345},
  number={6194},
  pages={1247125},
  year={2014},
  publisher={American Association for the Advancement of Science}
}

@article{clevers2016modeling,
  title={Modeling development and disease with organoids},
  author={Clevers, Hans},
  journal={Cell},
  volume={165},
  number={7},
  pages={1586--1597},
  year={2016},
  publisher={Elsevier}
}

@article{sigurbjornsdottir2014molecular,
  title={Molecular mechanisms of de novo lumen formation},
  author={Sigurbj{\"o}rnsd{\'o}ttir, Sara and Mathew, Renjith and Leptin, Maria},
  journal={Nature Reviews Molecular Cell Biology},
  volume={15},
  number={10},
  pages={665--676},
  year={2014},
  publisher={Nature Publishing Group UK London}
}

@article{lubarsky2003tube,
  title={Tube morphogenesis: making and shaping biological tubes},
  author={Lubarsky, Barry and Krasnow, Mark A},
  journal={Cell},
  volume={112},
  number={1},
  pages={19--28},
  year={2003},
  publisher={Elsevier}
}

@article{villasenor2010epithelial,
  title={Epithelial dynamics of pancreatic branching morphogenesis},
  author={Villasenor, Alethia and Chong, Diana C and Henkemeyer, Mark and Cleaver, Ondine},
  journal={Development},
  volume={137},
  number={24},
  pages={4295--4305},
  year={2010},
  publisher={Company of Biologists}
}

@article{bastidas2017cellular,
  title={Cellular and molecular mechanisms coordinating pancreas development},
  author={Bastidas-Ponce, Aim{\'e}e and Scheibner, Katharina and Lickert, Heiko and Bakhti, Mostafa},
  journal={Development},
  volume={144},
  number={16},
  pages={2873--2888},
  year={2017},
  publisher={The Company of Biologists Ltd}
}

@article{greggio2013artificial,
  title={Artificial three-dimensional niches deconstruct pancreas development in vitro},
  author={Greggio, Chiara and De Franceschi, Filippo and Figueiredo-Larsen, Manuel and Gobaa, Samy and Ranga, Adrian and Semb, Henrik and Lutolf, Matthias and Grapin-Botton, Anne},
  journal={Development},
  volume={140},
  number={21},
  pages={4452--4462},
  year={2013},
  publisher={Company of Biologists}
}

@article{torres2021tissue,
  title={Tissue hydraulics: Physics of lumen formation and interaction},
  author={Torres-S{\'a}nchez, Alejandro and Winter, Max Kerr and Salbreux, Guillaume},
  journal={Cells \& Development},
  volume={168},
  pages={203724},
  year={2021},
  publisher={Elsevier}
}

@article{chan2020integration,
  title={Integration of luminal pressure and signalling in tissue self-organization},
  author={Chan, Chii J and Hiiragi, Takashi},
  journal={Development},
  volume={147},
  number={5},
  pages={dev181297},
  year={2020},
  publisher={The Company of Biologists Ltd}
}

@article{nardini2021topological,
  title={Topological data analysis distinguishes parameter regimes in the Anderson-Chaplain model of angiogenesis},
  author={Nardini, John T and Stolz, Bernadette J and Flores, Kevin B and Harrington, Heather A and Byrne, Helen M},
  journal={PLOS Computational Biology},
  volume={17},
  number={6},
  pages={e1009094},
  year={2021},
  publisher={Public Library of Science San Francisco, CA USA}
}

@article{mcdonald2026topological,
  title={Topological model selection: a case-study in tumour-induced angiogenesis},
  author={McDonald, Robert A and Byrne, Helen M and Harrington, Heather A and Thorne, Thomas and Stolz, Bernadette J},
  journal={Bioinformatics},
  volume={42},
  number={3},
  pages={btag065},
  year={2026},
  publisher={Oxford University Press}
}

@article{turner2014persistent,
  title={Persistent homology transform for modeling shapes and surfaces},
  author={Turner, Katharine and Mukherjee, Sayan and Boyer, Doug M},
  journal={Information and Inference: A Journal of the IMA},
  volume={3},
  number={4},
  pages={310--344},
  year={2014},
  publisher={Oxford University Press}
}

@book{kendall1999shape,
  title     = {Shape and Shape Theory},
  author    = {Kendall, David G. and Barden, Dennis and Carne, Timothy K. and Le, Huiling},
  year      = {1999},
  publisher = {John Wiley \& Sons},
  address   = {Chichester, England},
  isbn      = {9780471970395},
}

@article{cooper2020chaste,
  title={Chaste: cancer, heart and soft tissue environment},
  author={Cooper, Fergus R and Baker, Ruth E and Bernabeu, Miguel O and Bordas, Rafel and Bowler, Louise and Bueno-Orovio, Alfonso and Byrne, Helen M and Carapella, Valentina and Cardone-Noott, Louie and Jonatha, Cooper and others},
  journal={Journal of Open Source Software},
  volume={5},
  number={47},
  pages={1848},
  year={2020}
}

@article{osborne2017comparing,
  title={Comparing individual-based approaches to modelling the self-organization of multicellular tissues},
  author={Osborne, James M and Fletcher, Alexander G and Pitt-Francis, Joe M and Maini, Philip K and Gavaghan, David J},
  journal={PLOS Computational Biology},
  volume={13},
  number={2},
  pages={e1005387},
  year={2017},
  publisher={Public Library of Science San Francisco, CA USA}
}

@article{van2015simulating,
  title={Simulating tissue mechanics with agent-based models: concepts, perspectives and some novel results},
  author={Van Liedekerke, Paul and Palm, MM and Jagiella, Nick and Drasdo, Dirk},
  journal={Computational Particle Mechanics},
  volume={2},
  number={4},
  pages={401--444},
  year={2015},
  publisher={Elsevier}
}

@article{graner1992simulation,
  title={Simulation of biological cell sorting using a two-dimensional extended Potts model},
  author={Graner, Fran{\c{c}}ois and Glazier, James A},
  journal={Physical Review Letters},
  volume={69},
  number={13},
  pages={2013},
  year={1992},
  publisher={APS}
}

@article{kursawe2018approximate,
  title={Approximate Bayesian computation reveals the importance of repeated measurements for parameterising cell-based models of growing tissues},
  author={Kursawe, Jochen and Baker, Ruth E and Fletcher, Alexander G},
  journal={Journal of Theoretical Biology},
  volume={443},
  pages={66--81},
  year={2018},
  publisher={Elsevier}
}

@article{kesavan2009cdc42,
  title={Cdc42-mediated tubulogenesis controls cell specification},
  author={Kesavan, Gokul and Sand, Fredrik Wolfhagen and Greiner, Thomas Uwe and Johansson, Jenny Kristina and Kobberup, Sune and Wu, Xunwei and Brakebusch, Cord and Semb, Henrik},
  journal={Cell},
  volume={139},
  number={4},
  pages={791--801},
  year={2009},
  publisher={Elsevier}
}

@inproceedings{larsen2017molecular,
  title={The molecular and morphogenetic basis of pancreas organogenesis},
  author={Larsen, Hjalte List and Grapin-Botton, Anne},
  booktitle={Seminars in cell \& developmental biology},
  volume={66},
  pages={51--68},
  year={2017},
  organization={Elsevier}
}

@misc {PPR:PPR811193,
	Title = {Generic comparison of lumen nucleation and fusion in epithelial organoids with and without hydrostatic pressure},
	Author = {Lu, Linjie and Fuji, Kana and Guyomar, Tristan and Alcheikh, Yara and Lieb, Michele and Andre, Marie and Tanida, Sakurako and Nonomura, Makiko and Hiraiwa, Tetsuya and Yennek, Siham and Petzold, Heike and Martin-Lemaitre, Cecilie and Grapin-Botton, Anne and Honigmann, Alf and Sano, Masaki and Riveline, Daniel},
	DOI = {10.1101/2024.02.20.581158},
	howpublished = {bioRxiv},
	Year = {2024},
	URL = {https://doi.org/10.1101/2024.02.20.581158},
}

@inproceedings{krull2019noise2void,
  title={Noise2void-learning denoising from single noisy images},
  author={Krull, Alexander and Buchholz, Tim-Oliver and Jug, Florian},
  booktitle={Proceedings of the IEEE/CVF conference on computer vision and pattern recognition},
  pages={2129--2137},
  year={2019}
}

@misc{pyclesperanto,
  author    = {Haase, Robert and Rajasekhar, Pradeep and Lambert, Talley and others},
  title     = {{clEsperanto/pyclesperanto\_prototype}, version 0.24.2},
  year      = {2023},
  publisher = {Zenodo},
  doi       = {10.5281/zenodo.10432619},
  url       = {https://doi.org/10.5281/zenodo.10432619}
}

@misc{apoc,
  author    = {Haase, Robert and Lee, Dohyeon and Doncila Pop, Draga and {\v{Z}}igutyt{\.{e}}, Laura},
  title     = {{napari-accelerated-pixel-and-object-classification} ({APOC}), version 0.14.1},
  year      = {2023},
  publisher = {Zenodo},
  doi       = {10.5281/zenodo.10071078},
  url       = {https://doi.org/10.5281/zenodo.10071078}
}

@article{prat2026dark,
  title={Dark Energy Survey Year 3 results: w CDM cosmology from simulation-based inference with persistent homology on the sphere},
  author={Prat, Judit and Gatti, M and Doux, C and Pranav, P and Chang, C and Jeffrey, N and Whiteway, L and Anbajagane, D and Sugiyama, S and Thomsen, A and others},
  journal={Monthly Notices of the Royal Astronomical Society},
  volume={545},
  number={3},
  pages={staf2152},
  year={2026},
  publisher={Oxford University Press}
}

@misc{li2025counting,
  title={Counting voids and filaments: Betti Curves as a Powerful Probe for Cosmology},
  author={Li, Jiayi and Zhao, Cheng},
  year={2025},
  howpublished={arXiv:2512.07236}
}

@misc{wenzel2025topologically,
  title={Topologically-based parameter inference for agent-based model selection from spatiotemporal cellular data},
  author={Wenzel, Alyssa R and Haughey, Patrick M and Nguyen, Kyle C and Nardini, John T and Haugh, Jason M and Flores, Kevin B},
  howpublished={bioRxiv},
  year={2025},
  url = {https://doi.org/10.1101/2025.06.13.659586}
}

@article{taylor2015topological,
  title={Topological data analysis of contagion maps for examining spreading processes on networks},
  author={Taylor, Dane and Klimm, Florian and Harrington, Heather A and Kram{\'a}r, Miroslav and Mischaikow, Konstantin and Porter, Mason A and Mucha, Peter J},
  journal={Nature Communications},
  volume={6},
  number={1},
  pages={7723},
  year={2015},
  publisher={Nature Publishing Group UK London}
}

@misc{liu2026multi,
  title={Multi-objective Bayesian inference in an agent-based model of zebrafish patterns via topological data analysis},
  author={Liu, Yue and Volkening, Alexandria},
  year={2026},
  howpublished={arXiv:2605.18685}
}

@article{zhou2026development,
  title={Development of a universal imaging “phenome” using shape, appearance and motion (SAM) features and the SAM Phenotype Observation Tool (SPOT)},
  author={Zhou, Felix Y and Norton-Steele, Adam and Marsh, Lewis and Byrne, Helen M and Harrington, Heather A and Lu, Xin},
  journal={Nature Communications},
  volume={17},
  number={1},
  pages={8409},
  year={2026},
  publisher={Nature Publishing Group UK London}
}

@article{zhou2026identifying,
  title={Identifying phenotype-genotype-function coupling in 3D organoid imaging using Shape, Appearance and Motion Phenotype Observation Tool (SPOT)},
  author={Zhou, Felix Y and Jacobs, Brittany-Amber and Norton-Steele, Adam and Han, Xiaoyue and Zhou, Linna and Carroll, Thomas M and Ruiz Puig, Carlos and Chadwick, Joseph and Qin, Xiao and Lisle, Richard and others},
  journal={Nature Communications},
  volume={17},
  number={1},
  pages={8410},
  year={2026},
  publisher={Nature Publishing Group UK London}
}

\end{document}